\documentclass[]{article}
\usepackage{preamble}
\usepackage{arxiv}
\usepackage[utf8]{inputenc} 
\usepackage[T1]{fontenc}    
\usepackage{url}            
\usepackage{booktabs}       
\usepackage{amsfonts}       
\usepackage{nicefrac}       
\usepackage{microtype}      
\usepackage{lipsum}		
\usepackage{graphicx}
\usepackage{natbib}
\usepackage{doi}
\usepackage{amsthm}

\title{Quantifying the HIV reservoir with dilution assays and deep viral sequencing}
\date{25 September 2023}	

\author{ \href{https://orcid.org/0000-0001-5380-2427}{\includegraphics[scale=0.06]{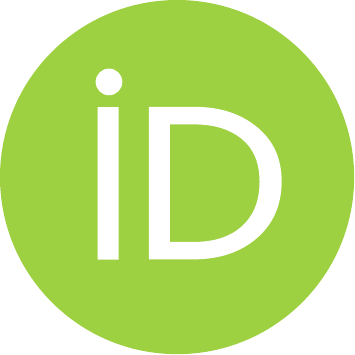}\hspace{1mm}Sarah C.~Lotspeich}\thanks{These authors contributed equally to the work.} \\
Department of Statistical Sciences\\Wake Forest University\\ 
Winston-Salem, North Carolina, U.S.A.\\ 
\And
\href{https://orcid.org/0000-0002-8870-5499}{\includegraphics[scale=0.06]{orcid.pdf}\hspace{1mm}Brian D.~Richardson$^\dagger$} \\
Department of Biostatistics\\
University of North Carolina at Chapel Hill\\
Chapel Hill, North Carolina, U.S.A. \\
\And
\href{https://orcid.org/0000-0002-9510-8326}{\includegraphics[scale=0.06]{orcid.pdf}\hspace{1mm}Pedro L.~Baldoni} \\
The Walter and Eliza Hall Institute of Medical Research\\ Parkville, Victoria, Australia \\
\And
Kimberly P.~Enders \\
Department of Biostatistics\\
University of North Carolina at Chapel Hill\\
Chapel Hill, North Carolina, U.S.A. \\
\And
\href{https://orcid.org/0000-0002-9106-4194}{\includegraphics[scale=0.06]{orcid.pdf}\hspace{1mm}
Michael G.~Hudgens} \\
Department of Biostatistics\\
University of North Carolina at Chapel Hill\\
Chapel Hill, North Carolina, U.S.A. \\
\texttt{mhudgens@email.unc.edu} \\
}

\renewcommand{\shorttitle}{Quantifying the HIV reservoir with dilution assays and deep viral sequencing}

\hypersetup{
pdftitle={Quantifying the HIV reservoir with dilution assays and deep viral sequencing},
pdfsubject={stat.AP, stat.TH},
pdfauthor={Sarah C.~Lotspeich, Brian D.~Richardson, Pedro L.~Baldoni, Kimberly P.~Enders, Michael G.~Hudgens},
pdfkeywords={Distinct viral lineages, infectious units per million, maximum likelihood estimation, missing data, Poisson distribution, serial limiting dilution assay.},
}

\begin{document}
\maketitle

\begin{abstract}
People living with HIV on antiretroviral therapy often have undetectable virus levels by standard assays, but ``latent'' HIV still persists in viral reservoirs. Eliminating these reservoirs is the goal of HIV cure research. The quantitative viral outgrowth assay (QVOA) is commonly used to estimate the reservoir size, i.e., the infectious units per million (IUPM) of HIV-persistent resting CD4+ T cells. A new variation of the QVOA, the Ultra Deep Sequencing Assay of the outgrowth virus (UDSA), was recently developed that further quantifies the number of viral lineages within a subset of infected wells. Performing the UDSA on a subset of wells provides additional information that can improve IUPM estimation. This paper considers statistical inference about the IUPM from combined dilution assay (QVOA) and deep viral sequencing (UDSA) data, even when some deep sequencing data are missing. Methods are proposed to accommodate assays with wells sequenced at multiple dilution levels and with imperfect sensitivity and specificity, and a novel bias-corrected estimator is included for small samples. The proposed methods are evaluated in a simulation study, applied to data from the University of North Carolina HIV Cure Center, and implemented in the open-source R package \texttt{SLDeepAssay}.
\end{abstract}


\section{Introduction}
\label{s:intro}

Modern antiretroviral therapy (ART) is a highly effective treatment for people living with HIV, often helping them achieve viral suppression (i.e., have a level of virus in their blood that is below the limit of detection of standard assays) and eliminating their risk of transmission to others. However, despite viral suppression, ``latent'' HIV-infected cells, which do not produce viral proteins and are not recognized by the immune system, will remain. These latently infected cells are commonly referred to as the HIV reservoir \citep{Ndung'u2019}. If a person living with HIV stops taking ART, these latently infected cells will result in viral rebound, sometimes in a matter of weeks \citep{LiEtAl2022}. Thus, the continued use of ART is necessary to maintain viral suppression, but there are costs and potential toxicities associated with lifelong use \citep{ChawlaEtAl2018}. Furthermore, as of 2021, only an estimated 75\% of the 38.4 million people living with HIV worldwide currently have access to treatment \citep{UNAIDS}, and it is unclear whether a feasible path towards 100\% treatment coverage exists. For these reasons, developing a cure for HIV that eliminates the latent viral reservoir and removes the need for ART is of high scientific and public health importance \citep{Ndung'u2019}. 

In HIV cure studies, a primary endpoint is the concentration of latent HIV-infected cells, often measured in infectious units per million cells (IUPM). This concentration is not directly measurable and is typically estimated through a serial limiting dilution (SLD) assay, wherein wells with known dilution levels (i.e, known numbers of cells) are tested for the presence of at least one cell with infectious virus. Repeating this process over multiple replicate wells (i.e., wells with the same dilution level) and at various dilution levels provides information for estimating the IUPM in the source population of cells (i.e., the person taking ART).

The quantitative viral outgrowth assay (QVOA) is one standard SLD assay for quantifying the HIV reservoir, as measured by the IUPM of resting CD4+ T cells. The QVOA tests wells for the presence of the HIV p24 antigen, an indicator that at least one cell within the well is HIV-infected. Various statistical methods have been proposed for drawing inference about the IUPM based on data from dilution assays like the QVOA. \citet{Myers1994} proposed a maximum likelihood estimator (MLE) of the IUPM, along with a corresponding exact confidence interval derived by inverting the likelihood ratio test. \citet{Trumble2017} proposed a bias-corrected MLE (BC-MLE), adapted from \citet{Hepworth&Watson2009}, that corrects for upward bias of the MLE in small samples. The open-source \texttt{SLDAssay} software package implements the methods described above.

The Ultra Deep Sequencing Assay of the outgrowth virus (UDSA), a variation of the QVOA, is a newer SLD assay for measuring the latent HIV reservoir that tests for the presence of distinct viral lineages (DVLs) in each well. Whereas the QVOA tests only for the presence of HIV in a given well, the UDSA provides additional information about the number of DVLs therein. Assuming that most latently infected cells are infected with at most one DVL, knowing the number of DVLs provides an improved lower bound (relative to the QVOA) for the number of infected cells in that well. Often, the QVOA is initially performed to identify the wells that are infected with at least one DVL (i.e., are positive), and then the UDSA is performed on a subsample of positive wells; this process leads to a missing data problem. \citet{Lee2017} proposed an MLE of the IUPM that incorporates partially observed additional information from the UDSA.

This paper justifies and extends existing methods to quantify the HIV reservoir from dilution assay and deep viral sequencing data. The \citet{Lee2017} estimator is shown to be consistent and asymptotically normal, and a bias-corrected MLE that accounts for the additional information from the UDSA is introduced. The possibility of the UDSA not detecting all DVLs in the source population is considered. Further, the MLE is extended to accommodate assays with multiple dilution levels, fully capturing all available information, and assumptions about the distribution of the assay data and the perfect sensitivity and specificity of the assays are relaxed. The proposed methods are compared with existing methods via simulation studies and an application to real assay data from the University of North Carolina (UNC) HIV Cure Center. The rest of the paper proceeds as follows. In Section~\ref{methods}, notation is defined, assumptions are given, and the proposed methods are introduced. Simulation studies are presented in Section~\ref{sims}, and data from the UNC HIV Cure Center are analyzed in Section~\ref{realdata}. Extensions of the proposed methods allowing for overdispersion and imperfect assays are presented in Section~\ref{methods:overdispersion-sens-spec}, and a brief discussion is given in Section~\ref{discuss}.

\section{Methods}\label{methods}
\label{s:model}

For simplicity, Sections~\ref{methods:model_notation}--\ref{methods:undetected} assume that only assay data from a single dilution level of one million cells per well are utilized. In Section~\ref{methods:multiple_dilutions}, the methods are extended to the multiple dilution level setting.

\subsection{Model and data}\label{methods:model_notation}

Following \citet{Myers1994} \citet{Trumble2017}, and \citet{Lee2017}, assume:
\begin{enumerate}
\item[(A1)] Cells are sampled randomly into $M$ wells from a larger source population,
\item[(A2)] For each DVL, infected cells are randomly distributed among the wells, and
\item[(A3)] The QVOA and UDSA have perfect sensitivity and specificity.
\end{enumerate}

Assay data are collected in two stages. \textit{Stage 1 (QVOA):} First, let $X_{j}$ be a latent variable denoting the number of cells in well $j$ that are infected with any DVL of HIV, $j \in \{1, \dots, M\}$. From the QVOA, indicator variables $W_{j}=\textrm{I}(X_j \geq 1)$ are observed in place of $X_{j}$, where $W_{j} = 1$ if well $j$ is positive and $W_{j} = 0$ otherwise. \textit{Stage 2 (UDSA):}
Let $n \in \{1, 2, \dots\}$ be the number of DVLs detected across the deep-sequenced wells. Note that $n$ is a random quantity and can be less than the number of DVLs existing in the source population. In Section~\ref{methods:undetected}, it is shown that, for the purposes of maximum likelihood estimation, it is sufficient to consider only the $n$ detected DVLs in the likelihood. Then, let $X_{ij}$ be a latent variable denoting the number of cells in well $j$ that are infected with observed DVL $i$ of HIV, $i \in \{1, \dots, n\}$. In practice, the indicator variables $Z_{ij} = \textrm{I}\left(X_{ij} \geq 1\right)$ are observed directly from the UDSA instead of the $X_{ij}$. For a given well $j$, let the vector $\pmb{Z}_j = (Z_{1j},\dots,Z_{nj})^\T$ contain indicators of whether each of the $n$ DVLs was detected therein. The random variables from Stages 1 and 2 are related via $W_{j} = \textrm{I}(\sum_{i=1}^{n}Z_{ij} \geq 1)$.

Let the vector $\pmb{X}_i=(X_{i1},\dots,X_{iM})^\T$ contain the numbers of cells infected with DVL $i$ in wells $1$ through $M$. Suppose that the components $X_{ij}$ are independent (A2) and Poisson distributed with rate $\lambda_i \geq 0$, where the DVL-specific rate parameter $\lambda_{i}$ represents the mean number of cells per well infected with DVL $i$. The counts of infected cells $\pmb{X}_i$ should be approximately Poisson distributed when the number of cells per well is large and $\lambda_i$ is small \citep{Myers1994, Trumble2017}. Because so few cells are latently infected and, of those that are, most are infected by only one DVL, the indicators for each DVL in well $j$, $\bZ_j = (Z_{1j},\dots,Z_{nj})^\T$, 
are approximately independent. Then, the indicators $W_j$ and $Z_{ij}$ follow Bernoulli distributions with $\Pr_{\blambda}(W_{j}=1) = 1 - \exp\left(-\sum_{i=1}^{n}\lambda_i\right)$ and $\Pr_{\lambda_i}(Z_{ij} = 1) = 1 - \exp\left(-\lambda_{i}\right)$, for $\blambda = (\lambda_1, \dots, \lambda_n)^\T$. Because it is assumed that there are one million cells per well, $\sum_{i=1}^{n}\lambda_i$ is the IUPM, which will be denoted by $\Lambda$.

Often, not all of the $M_P$ positive wells are sequenced with the UDSA, which introduces missingness. Let $R_{j}$ be a complete data indicator for well $j$, defined such that $R_{j} = 1$ if the well has complete data and $R_{j} = 0$ otherwise. Complete data are available from the $m$ positive wells with the additional UDSA information and from the $M_{N}$ negative wells. (No data are missing from the negative wells because, under (A3), negative QVOA results imply that there are zero DVLs in the negative wells.) Thus, the number of wells with complete data is $\sum_{j=1}^{M}R_j = m + M_{N}$.

\subsection{Likelihood construction}\label{subsec:likelihood}

\begin{figure}
    \centering
    \includegraphics[width=16cm]{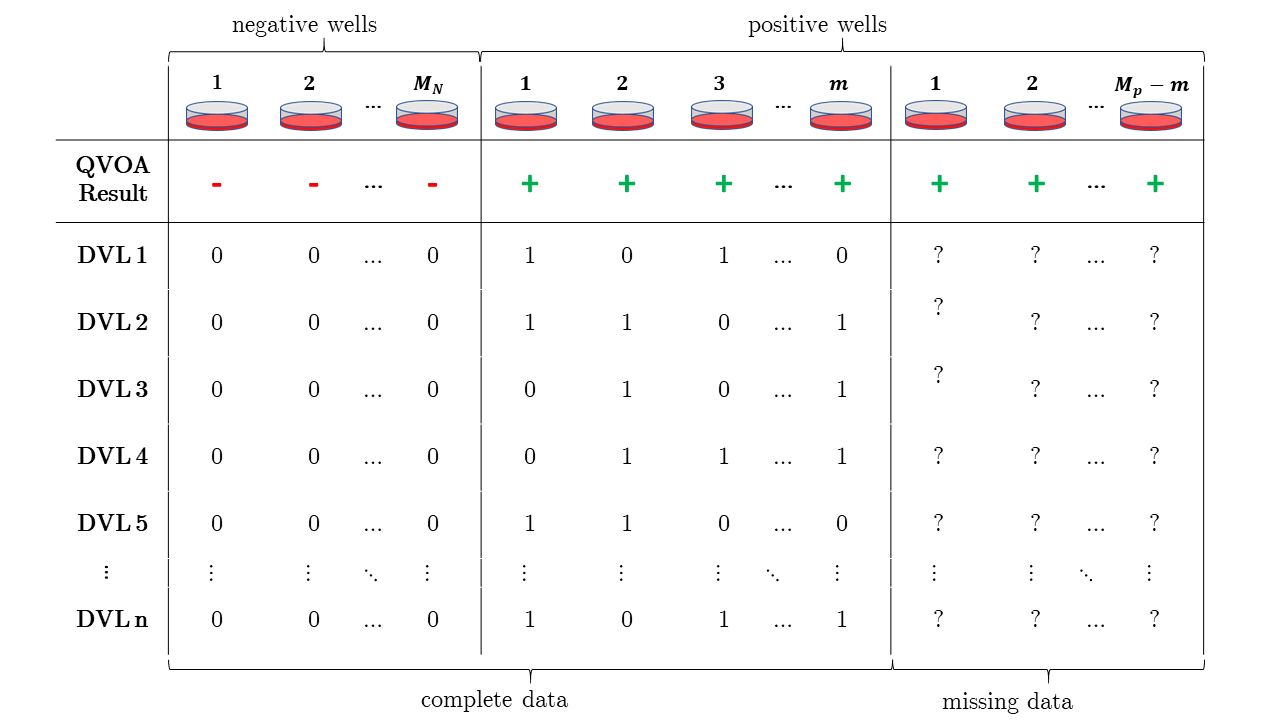}
    \caption{Illustration of the data collection scheme from the QVOA and UDSA at a single dilution level}
    \label{fig:fig1}
\end{figure}

All wells are initially tested for the presence of infectious virus using the QVOA, so the Stage 1 variables $\bW = (W_{1},\dots,W_{M})^\T$ are fully observed. However, since only a subset of the positive wells undergoes the UDSA in Stage 2, $\bZ = (\bZ_{1},\dots,\bZ_{M})$ will have missing data for the $M_P - m$ unsequenced positive wells. Based on this data collection scheme, illustrated in Figure~\ref{fig:fig1}, there are three types of well-level observations to consider: 
\begin{enumerate}
    \item[(T1)] A negative well ($R_{j} = 1, W_{j} = 0, \bZ_{j} = \pmb{0}$), 
    \item[(T2)] A positive well that was deep sequenced ($R_{j} = 1, W_{j} = 1, \bZ_{j} = \pmb{z}_j$), and 
    \item[(T3)] A positive well that was not deep sequenced ($R_{j} = 0, W_{j} = 1, \bZ_{j} = \pmb{?}$).
\end{enumerate}
For the two positive well types, at least one element in the $\bZ_j$ vector equals one since the well has to be positive for at least one DVL.

Incorporating all available information on all wells, the observed-data likelihood function is proportional to
\begin{align}
L(\blambda|\bW,\bZ,\bR) &= \prod_{j=1}^{M}{\Pr}_{\blambda}(W_j,\bZ_{j})^{R_j}{\Pr}_{\blambda}(W_{j})^{(1-R_j)}, \nonumber
\end{align}
where $\Pr_{\blambda}(W_j,\bZ_{j})$ is the joint probability mass function (PMF) of $(W_j, \bZ_j)$ and $\Pr_{\blambda}(W_j)$ is the marginal PMF of $W_j$. Because wells are selected for deep sequencing based only on the fully observed QVOA results $\bW$, the UDSA results $\bZ$ are missing at random (MAR) for the unsequenced wells \citep{Little&Rubin2002}. Therefore, the distribution of $\pmb{R}$ can be omitted from the likelihood for $\blambda$. Under the assumption of perfect QVOA sensitivity and specificity (A3), and since $W_j$ is completely determined by $\bZ_j$, $\Pr_{\blambda}(W_j|\bZ_{j}) = 1$, and it follows that $\Pr_{\blambda}(W_j,\bZ_{j}) = \Pr_{\blambda}(W_j|\bZ_{j})\Pr_{\blambda}(\bZ_{j}) = \Pr_{\blambda}(\bZ_{j})$. Assuming independence between the counts of infectious cells for the DVLs, $\Pr_{\blambda}(\bZ_{j}) = \prod_{i=1}^{n} \Pr_{\lambda_i}(Z_{ij})$. Thus, 
\begin{align*}
L(\blambda|\bW,\bZ,\bR) &= \prod_{j=1}^{M}\left[\prod_{i=1}^{n}\left\{1-\exp(-\lambda_i)\right\}^{Z_{ij}}\exp(-\lambda_i)^{(1-Z_{ij})}\right]^{R_j}\left\{1 - \exp\left(-\Lambda\right)\right\}^{(1-R_j)}, 
\end{align*}
which simplifies to
\begin{align*}
&\left[\prod_{i=1}^{n}\left\{1-\exp(-\lambda_i)\right\}^{\sum_{j=1}^{M}Z_{ij}R_{j}}\exp(-\lambda_i)^{\sum_{j=1}^{M}(1-Z_{ij})R_{j}}\right] \left\{1 - \exp\left(-\Lambda\right)\right\}^{\sum_{j=1}^{M}(1-R_j)}. 
\end{align*}

For ease of notation, let $Y_i = \sum_{j=1}^{M}Z_{ij}R_{j}$ and $\bY = (Y_1, \dots, Y_n)^\T$. It follows from the two-stage data collection procedure (Section~\ref{methods:model_notation}) that $\sum_{j=1}^{M}(1-Z_{ij})R_{j} = (M_N + m) - Y_i$ and $ \sum_{j=1}^{M}(1-R_j) = M - (M_N + m).$
Therefore, ($M_N, \bY$) are sufficient statistics for $\blambda$ and $L(\blambda|\bW,\bZ,\bR)$ 
can be rewritten as
\begin{align}
L(\blambda|M_N, \bY) &= \left[\prod_{i=1}^{n}\left\{1-\exp(-\lambda_i)\right\}^{Y_i}\exp(-\lambda_i)^{M_N + m - Y_i}\right]\left\{1 - \exp\left(-\Lambda\right)\right\}^{M - (M_N + m)}.
\label{likelihood_simplify}
\end{align}

\subsection{Maximum likelihood estimation}\label{methods:mle}

The MLE of the DVL-specific rate parameters, denoted $\widehat{\blambda} = (\widehat{\lambda}_1, \dots, \widehat{\lambda}_n)^{\T}$, is found by maximizing the observed-data log-likelihood $\log\{L(\blambda| M_N, \bY)\}$ based on \eqref{likelihood_simplify} with respect to $\blambda$, under the constraint that Poisson rates must be non-negative. When a subset of the positive wells are sequenced, the MLE $\blambdahat$ does not appear to have a closed-form solution but can be obtained numerically (see Web Appendix A for details). Analytical solutions do exist in two special cases: (i) when no positive wells are deep-sequenced or (ii) when all positive wells are deep-sequenced; see Web Appendices A.1 and A.2, respectively. There are also two extreme assay scenarios to consider. First, if all wells were QVOA negative, then no deep sequencing would be done, simplifying the likelihood and leading to an MLE for the IUPM of $\Lambdahat = 0$. Second, if (i) all wells were QVOA positive and (ii) all sequenced wells were UDSA positive for a single DVL, the MLE for the IUPM would be $\Lambdahat = \infty$.

Assuming that the UDSA data are indeed MAR and under suitable regularity conditions, the MLE $\widehat{\blambda}$ will be consistent for the true values $\blambda$ and asymptotically normally distributed \citep{Little&Rubin2002}. That is, $\sqrt{M}(\blambdahat - \blambda) \rightsquigarrow \mathcal{N}_n(\pmb{0}, \pmb{\Sigma})$, where $\rightsquigarrow$ denotes convergence in distribution and $\mathcal{N}_n(\pmb{0}, \pmb{\Sigma})$ is an $n$-variate normal distribution with mean vector $\pmb{0}$ and covariance matrix $\pmb{\Sigma}$. By the invariance property of MLEs, it follows that the MLE for the IUPM $\Lambda = \sum_{i=1}^{n}\lambda_i$ is $\Lambdahat=\sum_{i=1}^{n}\widehat{\lambda}_{i}$. Moreover, by the continuous mapping theorem and the delta method, $\Lambdahat$ is a consistent and asymptotically normal estimator of $\Lambda$. 

The asymptotic covariance matrix of $\blambdahat$ is given by the inverse of the Fisher information matrix, i.e.,  $\pmb{\Sigma} = \mathcal{I}(\blambda)^{-1} = \textrm{E}\left\{-\partial ^ 2 l(\blambda|M_N, \bY) /\partial \blambda \partial \blambda ^T \right\}^{-1}$, which can be consistently estimated by $\widehat{\pmb{\Sigma}} = \mathcal{I}(\blambda)^{-1}|_{\blambda=\blambdahat}$. A derivation of $\widehat{\pmb{\Sigma}}$ is given in Web Appendix B. Further, the standard error of the IUPM estimator $\Lambdahat$ can be estimated by $\widehat{\mathrm{SE}}(\Lambdahat) =  (\Sigma_{i=1}^n\Sigma_{j=1}^n \widehat{\Sigma}_{i,j})^{1/2}$, where $\widehat{\Sigma}_{i,j}$ denotes the $(i, j)$th element of $\widehat{\pmb{\Sigma}}$. By the delta method, a $100(1 - \alpha)\%$ Wald  confidence interval for the log IUPM $\log(\Lambda)$ has endpoints $\log(\Lambdahat) \pm z_{\alpha/2} \widehat{\mathrm{SE}}(\Lambdahat) / \Lambdahat$, where $z_{\alpha/2}$ denotes the $(1 - \alpha/2)$th percentile of the standard normal distribution. Exponentiating these endpoints gives a strictly positive confidence interval for $\Lambda$.

\subsection{Bias correction for small samples}
\label{methods:bcmle}

In SLD assay settings, the MLE will be upwardly biased with a small number of replicate wells $M$, such that $\Lambdahat$ tends to overestimate the size of a person's latent HIV reservoir \citep{Trumble2017}. A bias-corrected MLE (BC-MLE) based on \citet{Hepworth&Watson2009} was proposed by \citet{Trumble2017}. However, the \citet{Trumble2017} bias correction is intended for a one-dimensional parameter estimator and therefore cannot be applied to $\blambdahat$. Instead, a bias-correction method for the multi-dimensional setting, developed by \citet{Hashemi2021}, is adapted here. The method involves subtracting a correction term from the MLE $\blambdahat$ to reduce the order of the bias from $\sO(M^{-1})$ to $\sO(M^{-2})$.

Following \citet{Hashemi2021}, the bias of the MLE $\blambdahat$ can be expressed as 
\begin{align}
\textrm{E}\left(\blambdahat - \blambda\right) = \pmb{\Sigma}~\bA\!\left(\blambda\right) \mathrm{vec}\!\left(\pmb{\Sigma}\right) + \sO(M^{-2}), \label{bias}
\end{align}
where $\bA\!\left(\blambda\right)  = \left[\bA_1\!\left(\blambda\right) , \dots, \bA_n\!\left(\blambda\right)\right]$ is the $n \times n^2$ matrix with $n \times n$ submatrices
\begin{align*}
\bA_i\!\left(\blambda\right) = \frac{\partial}{\partial \lambda_i}\sI(\blambda) - \frac{1}{2} \textrm{E}\left\{ \frac{\partial^3}{\partial \blambda \partial \blambda ^T \partial \lambda_i}l(\blambda|M_N,\bY) \right\}
\end{align*}
and $\mathrm{vec}\!\left(\pmb{\Sigma}\right)$ denotes the $n^2 \times 1$ column vector obtained by stacking the columns of $\pmb{\Sigma}$. The components of submatrix $\bA_i\!\left(\blambda\right)$ are derived in Web Appendix C. Equation~\eqref{bias} motivates the BC-MLE for the DVL-specific rate parameters: $\pmb{\lambdahat}^* =  (\lambdahat_1^*, \dots, \lambdahat_n^*)^\T = \blambdahat - B(\blambdahat)$, where $B(\blambdahat) = \widehat{\pmb{\Sigma}} \bA(\blambdahat)\mathrm{vec}(\widehat{\pmb{\Sigma}})$. Finally, the BC-MLE for the IUPM is $\Lambdahat^* = \sum_{i=1}^n \lambdahat_{i}^*$.

Conveniently, the MLE and the BC-MLE have the same asymptotic distribution. To see this, note that the bias correction term $B(\blambdahat) = \sO_p(M^{-1})$, i.e., $MB\!\left(\blambdahat\right)\!$ is bounded in probability, so $\sqrt{M}B(\blambdahat)$ converges in probability to zero. Then, using Slutsky's theorem,
\begin{align*}
    \sqrt{M}\!\left(\pmb{\lambdahat}^* - \blambda\right) = \sqrt{M}\left[\left\{\blambdahat - B\!\left(\blambdahat\right)\right\} - \blambda\right] = \left\{\sqrt{M}\left(\blambdahat - \blambda\right) - \sqrt{M}B\!\left(\blambdahat\right)\right\} \rightsquigarrow \mathcal{N}_n(\pmb{0}, \pmb{\Sigma}),
\end{align*}
i.e., $\pmb{\lambdahat}^*$ is also a consistent and asymptotically normal estimator of $\blambda$. Moreover, the asymptotic covariance of $\pmb{\lambdahat}^*$ can be consistently estimated by $\widehat{\pmb{\Sigma}}$. However, by construction, $\pmb{\lambdahat}^*$ will tend to have have smaller bias than $\blambdahat$ for a small number of replicate wells.

\subsection{Estimation with undetected viral lineages}\label{methods:undetected}

When a person living with HIV is tested with the UDSA, only a very small subset of their CD4+ T cells are obtained (typically by leukapheresis). The person may have additional DVLs in their population of CD4+ T cells that were not present in the subset of cells sampled, in which case the UDSA would not detect these additional DVLs, even with perfect sensitivity and specificity. Below it is shown that the proposed IUPM estimator can still be viewed as an MLE, even in the presence of undetected viral lineages.

Suppose that there are $n'$ DVLs present in an individual living with HIV, $n$ of which are detected by the UDSA, $n' \in \{n + 1, n+2, n+3, \dots\}$. This leaves $n' - n$ undetected viral lineages, with corresponding counts of infected cells $\bY' = (Y_{n+1},\dots,Y_{n'})^{\T}$ that are independent and Poisson distributed with rates $\lambda_{i'}$, $i' \in \{n+1, \dots, n'\}$. With some abuse of notation, let $Y_{0} = Y_{n+1} + \dots + Y_{n'}$ denote the number of wells infected with any of the undetected DVLs. Then, $Y_{0}$ also has a Poisson distribution with rate $\lambda_{0} = \lambda_{n+1} + \dots + \lambda_{n'}$, and, since no wells are infected with these lineages, $Y_{0}=0$ is observed. Thus, the augmented likelihood function accounting for all DVLs (detected and undetected) can be written as 
\begin{align}
& L'(\blambda'|M_N, \bY, Y_{0}) \nonumber \\
&= \left[\prod_{i=0}^{n}\left\{1-\exp(-\lambda_i)\right\}^{Y_{i}}\exp(-\lambda_i)^{(M_N+m-Y_i)}\right]\left\{1 - \exp\left(-\sum_{i=0}^{n}\lambda_i\right)\right\}^{(M-M_{N}-m)},
\label{likelihood_und}
\end{align}
where $\blambda' = (\lambda_0,\blambda^\T)^{\T}$. Given that DVLs $n+1, \dots, n'$ are undetected, a reasonable heuristic estimate for the rate of their sum $\lambda_{0}$ is zero. In fact, it is proven in the Appendix that the MLE for $\lambda_{0}$ \textit{is} zero. That is, the vector $\widehat{\blambda'}$ that maximizes \eqref{likelihood_und} necessarily satisfies $\lambdahat_{0} = 0$. When estimating $\Lambda$, this implies that using the sum of the $n$-dimensional MLE $\blambdahat$ from the original likelihood in \eqref{likelihood_simplify} is equivalent to using the sum of the $(n+1)$-dimensional MLE $\widehat{\blambda'}$ from the augmented likelihood in \eqref{likelihood_und}. In other words, summing the DVL-specific MLEs for the detected DVLs gives the MLE for the sum of \textit{all the} DVL-specific rate parameters, detected or not.

\subsection{Incorporating multiple dilution levels}
\label{methods:multiple_dilutions}

So far, it has been assumed that the assay was conducted at a single dilution level, with each replicate well containing one million ($10^6$) cells. In practice, dilution levels other than one million cells per well may be used. Moreover, multiple dilution levels are often tested with the QVOA to pinpoint one or more appropriate dilution levels for the UDSA (i.e., dilution levels with sufficient positive wells). The methods from Sections~\ref{methods:model_notation}--\ref{methods:undetected} are now adapted to handle these two cases.

First, consider the setting where an assay is done at a single dilution level, but each replicate well contains $u \times 10^6$ cells for some $u>0$. Continue to let $\lambda_i$ be the mean count of cells per well infected with DVL $i$, $i \in \{1, \dots, n\}$. Now, let $\tau_i$ be the mean count of cells \textit{per million} infected with DVL $i$, and denote by $\btau = (\tau_1, \dots, \tau_n)^{\T}$ the vector of DVL-specific IUPMs for all $n$ DVLs. If $u=1$, as assumed in previous sections, then $\blambda = \btau$ and $\Lambda = T$. More generally, for a dilution level of $u$, the DVL-specific IUPMs and mean counts per cell are related through $\btau = \blambda/u$. Using this relationship, 
\eqref{likelihood_simplify} can be rewritten as a function of $\btau$ by substituting $\blambda = u\btau$ and defining $\widetilde{L}(\btau|M_N, \bY, u) = L(u\btau|M_N, \bY)$. Then, the MLE for $\btau$ is the value $\btauhat$ that maximizes $\widetilde{L}(\btau|M_N, \bY, u)$ over the parameter space $[0,\infty)^n$, and, from it, the IUPM is estimated as $\Tauhat = \sum_{i=1}^n\tauhat_i$.

Now, consider the second case where assay data $(M_{N}^{(d)}, \bY^{(d)}, u^{(d)})$, $d \in \{1, \dots, D\}$, are available from replicate wells at $D$ distinct dilution levels, $D \in \{1, 2, \dots \}$. Let $M_{N}^{(d)}, \bY^{(d)}$, and $u^{(d)}$ denote the number of negative wells, vector of summarized UDSA results, and dilution level, respectively, for the $d$th dilution. Assume independence between replicate wells and across dilution levels; this is the natural extension of assumption (A1) to the multiple dilution level setting. Then, the joint likelihood given data from all $D$ dilution levels is proportional to the product of the individual likelihoods given data from each dilution level: 
\begin{align}
\widetilde{L}(\btau|\pmb{M_N}, \bY, \pmb{u}) = \prod_{d=1}^D {\widetilde{L}}\left(\btau|M_{N}^{(d)}, \bY^{(d)}, u^{(d)}\right),
\label{joint-likelihood}
\end{align}
where $\pmb{M_N} = (M_{N}^{(1)},\dots,M_{N}^{(D)})^\T$, $\bY = (\bY^{(1)},\dots,\bY^{(D)})$, and $\pmb{u} = (u^{(1)}, \dots, u^{(D)})^\T$. The MLE for the vector of DVL-specific IUPMs is the value $\btauhat$ that maximizes  (\ref{joint-likelihood}), and the corresponding MLE for the IUPM can be calculated as their sum. 

The MLE $\Tauhat$ from 
\eqref{joint-likelihood} is once again consistent for the true IUPM $T$ and asymptotically normally distributed with asymptotic variance $\sum_{i=1}^n\sum_{j=1}^n\widetilde{\Sigma}_{ij}$, where $\widetilde{\pmb{\Sigma}} = \widetilde{\mathcal{I}}({\btau})^{-1} = \textrm{E}[-\partial ^ 2 \ln\left\{\widetilde{L}(\btau|\pmb{M_N}, \bY, \pmb{u})\right\} /\partial \btau \partial \btau ^T]^{-1}$. The same bias correction method introduced in Section~\ref{methods:bcmle} can be applied to the multiple dilution level setting. Details on estimating $\widetilde{\pmb{\Sigma}}$ and computing the bias correction term in this setting are given in Web Appendix D. Note that the likelihood from \citet{Myers1994}, which uses QVOA data only, is a special case of 
\eqref{joint-likelihood} where none of the positive wells are deep sequenced (i.e., $m=0$). Thus, \eqref{joint-likelihood} can be used even when deep sequencing is not done.

\section{Simulations}\label{sims}
\label{s:sim}

Simulation studies were performed to assess the proposed methods. Various settings were considered, inspired by real-world dilution assay studies with a single (Section~\ref{sims:single}) or multiple (Section~\ref{sims:multiple}) dilution levels. In addition to demonstrating the methods' validity, these simulations illustrate the notable efficiency gains from incorporating deep viral sequencing.

\subsection{Simulations with a single dilution level}\label{sims:single}

Data for a single dilution assay were simulated as follows. First, full results from the UDSA were generated as the DVL-specific infection indicators $Z_{ij}$ for all wells, $j \in \{1, \dots, M\}$, and DVLs, $i \in \{1, \dots, n'\}$, from independent Bernoulli distributions with $\Pr(Z_{ij} = 1) = 1 - \exp(-\lambda_i)$. Results from the QVOA were then calculated as $W_{j} = \textrm{I}(\sum_{i=1}^{n}Z_{ij} \geq 1)$ for all wells. The number of wells to undergo the UDSA was computed as $m = \nint{qM_P}$, where $q$ was the fixed proportion of positive wells that undergo the UDSA and $\nint{\cdot}$ denotes the nearest integer function. Based on $q$, a random sample of $M_P - m$ positive wells were set to be missing their $Z_{ij}$ information for all DVLs.

The simulation studies utilized a fully factorial design by considering all possible combinations of $M = 12, 24$, or $32$ replicate wells; $n' = 6, 12$, or $18$ DVLs; proportions $q = 1, 0.75,$ or $0.5$ of positive wells that underwent the UDSA; and IUPM $T = 0.5$ or $1$. These choices of parameters, which were motivated by the real data used in Section~\ref{realdata}, led to 54 unique simulation settings defined by ($M, n', q, T$). For simplicity, the single dilution level was chosen to be $u = 1$ million cells per well, so $T = \Lambda$. In addition, two allocations of the IUPM $T$ across the $n'$ DVLs were considered: (i) \textit{constant rate}, i.e., the same IUPM for all DVLs such that $\tau_i = T / n'$ for $i \in \{1, \dots, n'\}$, and (ii) \textit{non-constant rate}, i.e., a larger IUPM for the last $n' / 2$ DVLs such that $\tau_i = T / (2n')$ for $i \in \{1, \dots, n'/2\}$ versus $\tau_i = 3T/(2n')$ for $i \in \{n'/2+1, \dots, n'\}$. Both allocations were applied for $T = 1$; for $T = 0.5$ only the constant rate scenario was considered. Two extreme results were possible: (i) all wells were negative, in which case the UDSA would not be done and the IUPM estimator would be zero, or (ii) all wells were positive and a particular DVL was detected in each deep sequenced well, in which case the IUPM estimator would be infinite. While (i) never happened, (ii) occurred in 45 out of \num{81000} simulations ($<0.1\%$); in these cases the simulated assay data were discarded and resimulated.

Four IUPM estimators were applied to each simulated assay: (i) MLE without UDSA, (ii) BC-MLE without UDSA, (iii) MLE with UDSA, and (iv) BC-MLE with UDSA. All estimators have been implemented in \texttt{R} packages, with (i) and (ii) in \texttt{SLDAssay} \citep{Trumble2017} and (iii) and (iv) in \texttt{SLDeepAssay} (newly developed to accompany this paper). Estimators (ii) and (iv) were expected to have smaller bias than (i) and (iii) in small samples, and estimators (iii) and (iv) were expected to be more precise than (i) and (ii) due to the added DVL information from the UDSA.

A number of metrics are reported for comparison of the four IUPM estimators, summarizing the \num{1000} data sets simulated for each setting. The relative bias (``bias'') was computed by dividing (i) the mean differences between the estimated and true IUPM across replications by (ii) the true IUPM. The average standard error (ASE) and empirical standard error (ESE) were computed as the empirical mean of the standard error estimator and the empirical standard deviation of the IUPM estimates, respectively. Finally, the empirical coverage probability (CP) for the 95\% confidence interval was computed as the proportion of simulations where the true IUPM fell between the lower and upper bounds of the interval.

Detailed results for a single dilution assay with IUPM of $T = 1$ and a constant rate of infected cells for all DVLs can be found in Table~\ref{Table1}. As expected, the two BC-MLEs had very little bias in all settings (both $\leq 4\%$). Meanwhile, the uncorrected MLEs saw bias as large as 9\%; this bias improved, though, as either (i) the number of wells $M$ increased or (ii) the proportion $q$ being deep sequenced increased (for the estimators with UDSA). Bias for all estimators was unchanged by an increasing number of DVLs $n'$.

\begin{table}
\centering
\caption{Simulation results with a single dilution level, assuming a constant rate of infected cells for all distinct viral lineages. The true IUPM in all settings was $T = 1$.}
\label{Table1}
\resizebox{\columnwidth}{!}{
\begin{threeparttable}
\begin{tabular}{cccrrrrrrrrrrrrrrrr}
\toprule
\multicolumn{3}{c}{\textbf{ }} & \multicolumn{8}{c}{\textbf{Without UDSA}} & \multicolumn{8}{c}{\textbf{With UDSA}} \\
\cmidrule(l{3pt}r{3pt}){4-11} \cmidrule(l{3pt}r{3pt}){12-19}
\multicolumn{3}{c}{\textbf{ }} & \multicolumn{4}{c}{\textbf{MLE}} & \multicolumn{4}{c}{\textbf{Bias-Corrected MLE}} & \multicolumn{4}{c}{\textbf{MLE}} & \multicolumn{4}{c}{\textbf{Bias-Corrected MLE}} \\
\cmidrule(l{3pt}r{3pt}){4-7} \cmidrule(l{3pt}r{3pt}){8-11} \cmidrule(l{3pt}r{3pt}){12-15} \cmidrule(l{3pt}r{3pt}){16-19}
\textbf{$\pmb{n'}$} & $\pmb{M}$& $\pmb{q}$ & \textbf{Bias} & \textbf{ASE} & \textbf{ESE} & \textbf{CP} & \textbf{Bias} & \textbf{ASE} & \textbf{ESE} & \textbf{CP} & \textbf{Bias} & \textbf{ASE} & \textbf{ESE} & \textbf{CP} & \textbf{Bias} & \textbf{ASE} & \textbf{ESE} & \textbf{CP}\\
\midrule
6 & 12 & 0.50 & $0.08$ & $0.41$ & $0.45$ & $0.95$ & $-0.02$ & $0.41$ & $0.37$ & $0.89$ & $0.06$ & $0.35$ & $0.36$ & $0.94$ & $-0.04$ & $0.35$ & $0.33$ & $0.98$\\
&&  0.75 & $0.08$ & $0.41$ & $0.45$ & $0.95$ & $-0.02$ & $0.41$ & $0.37$ & $0.89$ & $0.05$ & $0.33$ & $0.34$ & $0.94$ & $-0.02$ & $0.33$ & $0.32$ & $0.97$\\
&&  1.00 & $0.08$ & $0.41$ & $0.45$ & $0.95$ & $-0.02$ & $0.41$ & $0.37$ & $0.89$ & $0.04$ & $0.31$ & $0.32$ & $0.94$ & $-0.01$ & $0.31$ & $0.30$ & $0.96$\\
\addlinespace
& 24 & 0.50 & $0.04$ & $0.28$ & $0.28$ & $0.97$ & $ 0.00$ & $0.28$ & $0.26$ & $0.94$ & $0.04$ & $0.24$ & $0.24$ & $0.95$ & $-0.01$ & $0.24$ & $0.23$ & $0.97$\\
&&  0.75 & $0.04$ & $0.28$ & $0.28$ & $0.97$ & $ 0.00$ & $0.28$ & $0.26$ & $0.94$ & $0.02$ & $0.23$ & $0.23$ & $0.95$ & $-0.01$ & $0.23$ & $0.22$ & $0.96$\\
&&  1.00 & $0.04$ & $0.28$ & $0.28$ & $0.97$ & $ 0.00$ & $0.28$ & $0.26$ & $0.94$ & $0.02$ & $0.22$ & $0.21$ & $0.96$ & $-0.01$ & $0.22$ & $0.20$ & $0.96$\\
\addlinespace
& 32 & 0.50 & $0.02$ & $0.24$ & $0.24$ & $0.96$ & $-0.01$ & $0.24$ & $0.23$ & $0.91$ & $0.02$ & $0.21$ & $0.20$ & $0.96$ & $-0.01$ & $0.21$ & $0.20$ & $0.97$\\
&&  0.75 & $0.02$ & $0.24$ & $0.24$ & $0.96$ & $-0.01$ & $0.24$ & $0.23$ & $0.91$ & $0.01$ & $0.20$ & $0.19$ & $0.96$ & $-0.01$ & $0.20$ & $0.19$ & $0.97$\\
&&  1.00 & $0.02$ & $0.24$ & $0.24$ & $0.96$ & $-0.01$ & $0.24$ & $0.23$ & $0.91$ & $0.01$ & $0.19$ & $0.18$ & $0.95$ & $-0.01$ & $0.19$ & $0.18$ & $0.96$\\
\addlinespace
12 & 12 & 0.50 & $0.08$ & $0.41$ & $0.44$ & $0.96$ & $-0.02$ & $0.41$ & $0.36$ & $0.89$ & $0.08$ & $0.35$ & $0.37$ & $0.94$ & $-0.02$ & $0.35$ & $0.34$ & $0.97$\\
&&  0.75 & $0.08$ & $0.41$ & $0.44$ & $0.96$ & $-0.02$ & $0.41$ & $0.36$ & $0.89$ & $0.06$ & $0.32$ & $0.34$ & $0.94$ & $ 0.00$ & $0.32$ & $0.32$ & $0.96$\\
&&  1.00 & $0.08$ & $0.41$ & $0.44$ & $0.96$ & $-0.02$ & $0.41$ & $0.36$ & $0.89$ & $0.04$ & $0.30$ & $0.31$ & $0.94$ & $-0.01$ & $0.30$ & $0.30$ & $0.96$\\
\addlinespace
& 24 & 0.50 & $0.04$ & $0.28$ & $0.30$ & $0.96$ & $ 0.00$ & $0.28$ & $0.28$ & $0.94$ & $0.03$ & $0.24$ & $0.24$ & $0.95$ & $-0.01$ & $0.24$ & $0.23$ & $0.96$\\
&&  0.75 & $0.04$ & $0.28$ & $0.30$ & $0.96$ & $ 0.00$ & $0.28$ & $0.28$ & $0.94$ & $0.02$ & $0.22$ & $0.22$ & $0.96$ & $-0.01$ & $0.22$ & $0.22$ & $0.96$\\
&&  1.00 & $0.04$ & $0.28$ & $0.30$ & $0.96$ & $ 0.00$ & $0.28$ & $0.28$ & $0.94$ & $0.02$ & $0.21$ & $0.21$ & $0.95$ & $ 0.00$ & $0.21$ & $0.20$ & $0.96$\\
\addlinespace
& 32 & 0.50 & $0.03$ & $0.24$ & $0.25$ & $0.96$ & $ 0.00$ & $0.24$ & $0.24$ & $0.91$ & $0.02$ & $0.21$ & $0.21$ & $0.95$ & $-0.01$ & $0.21$ & $0.20$ & $0.95$\\
&&  0.75 & $0.03$ & $0.24$ & $0.25$ & $0.96$ & $ 0.00$ & $0.24$ & $0.24$ & $0.91$ & $0.02$ & $0.19$ & $0.19$ & $0.95$ & $ 0.00$ & $0.19$ & $0.19$ & $0.95$\\
&&  1.00 & $0.03$ & $0.24$ & $0.25$ & $0.96$ & $ 0.00$ & $0.24$ & $0.24$ & $0.91$ & $0.01$ & $0.18$ & $0.18$ & $0.95$ & $ 0.00$ & $0.18$ & $0.18$ & $0.95$\\
\addlinespace
18 & 12 & 0.50 & $0.09$ & $0.41$ & $0.42$ & $0.97$ & $-0.01$ & $0.41$ & $0.35$ & $0.90$ & $0.08$ & $0.35$ & $0.36$ & $0.94$ & $-0.01$ & $0.35$ & $0.33$ & $0.98$\\
&&  0.75 & $0.09$ & $0.41$ & $0.42$ & $0.97$ & $-0.01$ & $0.41$ & $0.35$ & $0.90$ & $0.06$ & $0.32$ & $0.32$ & $0.95$ & $ 0.00$ & $0.32$ & $0.30$ & $0.97$\\
&&  1.00 & $0.09$ & $0.41$ & $0.42$ & $0.97$ & $-0.01$ & $0.41$ & $0.35$ & $0.90$ & $0.05$ & $0.30$ & $0.30$ & $0.94$ & $ 0.00$ & $0.30$ & $0.29$ & $0.97$\\
\addlinespace
& 24 & 0.50 & $0.04$ & $0.28$ & $0.30$ & $0.95$ & $ 0.00$ & $0.28$ & $0.28$ & $0.94$ & $0.04$ & $0.24$ & $0.24$ & $0.94$ & $ 0.00$ & $0.24$ & $0.23$ & $0.96$\\
&&  0.75 & $0.04$ & $0.28$ & $0.30$ & $0.95$ & $ 0.00$ & $0.28$ & $0.28$ & $0.94$ & $0.03$ & $0.22$ & $0.22$ & $0.94$ & $ 0.00$ & $0.22$ & $0.22$ & $0.96$\\
&&  1.00 & $0.04$ & $0.28$ & $0.30$ & $0.95$ & $ 0.00$ & $0.28$ & $0.28$ & $0.94$ & $0.02$ & $0.21$ & $0.21$ & $0.95$ & $ 0.00$ & $0.21$ & $0.20$ & $0.96$\\
\addlinespace
& 32 & 0.50 & $0.03$ & $0.24$ & $0.25$ & $0.96$ & $ 0.00$ & $0.24$ & $0.23$ & $0.92$ & $0.03$ & $0.20$ & $0.21$ & $0.96$ & $ 0.00$ & $0.20$ & $0.20$ & $0.96$\\
&&  0.75 & $0.03$ & $0.24$ & $0.25$ & $0.96$ & $ 0.00$ & $0.24$ & $0.23$ & $0.92$ & $0.02$ & $0.19$ & $0.19$ & $0.95$ & $ 0.00$ & $0.19$ & $0.19$ & $0.96$\\
&&  1.00 & $0.03$ & $0.24$ & $0.25$ & $0.96$ & $ 0.00$ & $0.24$ & $0.23$ & $0.92$ & $0.02$ & $0.18$ & $0.18$ & $0.95$ & $ 0.00$ & $0.18$ & $0.18$ & $0.96$\\
\bottomrule
\end{tabular}
\begin{tablenotes}[flushleft]
\item{\em Note:} \textbf{Bias} and \textbf{ESE} are, respectively, the empirical relative bias and standard error of the IUPM estimator; \textbf{ASE} is the average of the standard error estimator; \textbf{CP} is the empirical coverage probability of the 95\% confidence interval for the IUPM. There were a total of 48 simulated assays out of \num{27000} ($0.2\%$) where the MLE and BC-MLE without UDSA were infinite that were excluded; all other entries are based on \num{1000} replicates.
\end{tablenotes}
\end{threeparttable}
}
\end{table}

Overall, the standard error estimators approximated the empirical standard errors well. For the MLE without UDSA, the ASE overestimated the ESE, but this was resolved as $M$ increased. Based on ASE or ESE, the variability of the estimators with UDSA decreased when there were more replicate wells (i.e., larger $M$) and, as expected, when the deep sequencing information was available for more wells (i.e., larger $q$). The empirical relative efficiency (RE) can be used to compare the statistical precision of the estimators with and without UDSA, computed as the ratio of the squared ESE of the estimator without UDSA to the squared ESE of the estimator with UDSA (i.e., the ratio of the empirical variances). In fact, the MLE and BC-MLE with UDSA were as much as $110\%$ (RE $= 2.10$) and $89\%$ (RE $= 1.89$) more efficient, respectively, than their counterparts without sequencing data. Despite needing to estimate more parameters when there were more DVLs, the variability of these estimators was stable as $n'$ increased. The confidence intervals for the BC-MLEs were sometimes conservative with the smallest number of $M = 12$ wells, but appeared reasonable for all other settings. Also, the over-coverage was slightly less severe for the estimators with UDSA than without. Otherwise, the confidence intervals achieved the appropriate coverage. 

Results with a non-constant rate of infected cells were nearly identical (Web Table S1). Aside from uniformly smaller standard errors, results with the smaller a IUPM of $T = 0.5$ were comparable to those discussed already (Web Table S2).  

\subsection{Simulations with multiple dilution levels}\label{sims:multiple}

Data for an assay at multiple dilution levels were simulated as in Section \ref{sims:single}, with a few modifications. For each scenario, three single dilution assay datasets were simulated, one for each of the $D = 3$ dilution levels, and then combined for analysis. The following parameters were held fixed: (i) the true IUPM $T=1$, (ii) the three dilution levels $\pmb{u} = (u_1, u_2, u_3) = (0.5, 1, 2)$ million cells per well, and (iii) the proportions of positive wells to be deep sequenced at the three dilution levels $\pmb{q} = (q_1, q_2, q_3) = (0, 0.5, 1)$. The simulation settings varied by the number of replicate wells per dilution level, $\pmb{M} = (M_1, M_2, M_3) = (6, 12, 18), (9, 18, 27)$, or $(12, 24, 36)$, and the number of DVLs, $n' = 6, 12$, or $18$. Again, the IUPM could be allocated across the $n'$ DVLs in a constant or non-constant way. No simulated data sets were discarded due to extreme scenarios with all wells being negative or positive (at all dilution levels). The same four estimators (MLE and BC-MLE, with and without UDSA) were applied to each simulated assay and compared with respect to bias, ASE, ESE, and CP.

Detailed results for the multiple dilution level assays with a constant rate of infected cells can be found in Table~\ref{sims-multiple}. 
In comparison to the single dilution simulations, the two uncorrected MLEs had relatively small bias ($\leq 5\%$ versus $\leq 9\%$). This improvement is likely due to the fact that these estimators incorporate more information from the multiple dilutions. Still, the two bias-corrected MLEs further reduced this bias to $\leq 1\%$ in all settings. The estimated standard errors were approximately consistent with the empirical ones, and the confidence intervals achieved near-nominal coverage, with empirical estimates between 93\% and 97\%. Across all settings, the estimators that used the UDSA had greater efficiency than those that did not, as reflected by the reductions in the ASE and ESE. As in the single dilution case, results with a non-constant rate of infected cells for the DVLs were nearly identical (Web Table S3).

\begin{table}
\centering
\caption{Simulation results with multiple dilution levels and a constant rate of infected cells for all distinct viral lineages. In all settings, the true IUPM was $T = 1$ and the proportions of positive wells that were deep sequenced at the three dilution levels were $\pmb{q} = (0, 0.5, 1)$.}
\label{sims-multiple}
\resizebox{\columnwidth}{!}{
\begin{threeparttable}
\begin{tabular}{ccccrrrrrrrrrrrrrr}
\toprule
\multicolumn{2}{c}{\textbf{ }} & \multicolumn{8}{c}{\textbf{Without UDSA}} & \multicolumn{8}{c}{\textbf{With UDSA}} \\
\cmidrule(l{3pt}r{3pt}){3-10} \cmidrule(l{3pt}r{3pt}){11-18}
\multicolumn{2}{c}{\textbf{ }} & \multicolumn{4}{c}{\textbf{MLE}} & \multicolumn{4}{c}{\textbf{Bias-Corrected MLE}} & \multicolumn{4}{c}{\textbf{MLE}} & \multicolumn{4}{c}{\textbf{Bias-Corrected MLE}} \\
\cmidrule(l{3pt}r{3pt}){3-6} \cmidrule(l{3pt}r{3pt}){7-10} \cmidrule(l{3pt}r{3pt}){11-14} \cmidrule(l{3pt}r{3pt}){15-18}
\textbf{$\pmb{n'}$} & $\pmb{M}$ & \textbf{Bias} & \textbf{ASE} & \textbf{ESE} & \textbf{CP} & \textbf{Bias} & \textbf{ASE} & \textbf{ESE} & \textbf{CP} & \textbf{Bias} & \textbf{ASE} & \textbf{ESE} & \textbf{CP} & \textbf{Bias} & \textbf{ASE} & \textbf{ESE} & \textbf{CP}\\
\midrule
6 & 6, 12, 18 & $0.04$ & $0.22$ & $0.24$ & $0.96$ & $ 0.00$ & $0.22$ & $0.22$ & $0.96$ & $0.02$ & $0.16$ & $0.15$ & $0.96$ & $-0.01$ & $0.16$ & $0.15$ & $0.97$\\
& 9, 18, 27 & $0.03$ & $0.18$ & $0.19$ & $0.95$ & $ 0.01$ & $0.18$ & $0.18$ & $0.95$ & $0.02$ & $0.13$ & $0.12$ & $0.96$ & $ 0.00$ & $0.13$ & $0.12$ & $0.97$\\
& 12, 24, 36 & $0.01$ & $0.16$ & $0.16$ & $0.95$ & $-0.01$ & $0.16$ & $0.15$ & $0.95$ & $0.01$ & $0.11$ & $0.11$ & $0.95$ & $ 0.00$ & $0.11$ & $0.11$ & $0.95$\\
\addlinespace
12 & 6, 12, 18 & $0.05$ & $0.22$ & $0.25$ & $0.96$ & $ 0.00$ & $0.22$ & $0.23$ & $0.96$ & $0.02$ & $0.15$ & $0.16$ & $0.94$ & $ 0.00$ & $0.15$ & $0.15$ & $0.96$\\
& 9, 18, 27 & $0.02$ & $0.18$ & $0.19$ & $0.95$ & $-0.01$ & $0.18$ & $0.18$ & $0.96$ & $0.01$ & $0.12$ & $0.12$ & $0.94$ & $ 0.00$ & $0.12$ & $0.12$ & $0.95$\\
& 12, 24, 36 & $0.02$ & $0.16$ & $0.16$ & $0.95$ & $ 0.00$ & $0.16$ & $0.16$ & $0.96$ & $0.01$ & $0.11$ & $0.10$ & $0.95$ & $ 0.00$ & $0.11$ & $0.10$ & $0.96$\\
\addlinespace
18 & 6, 12, 18 & $0.05$ & $0.22$ & $0.25$ & $0.94$ & $ 0.00$ & $0.22$ & $0.23$ & $0.95$ & $0.02$ & $0.15$ & $0.16$ & $0.93$ & $ 0.00$ & $0.15$ & $0.15$ & $0.95$\\
& 9, 18, 27 & $0.02$ & $0.18$ & $0.19$ & $0.95$ & $ 0.00$ & $0.18$ & $0.18$ & $0.96$ & $0.01$ & $0.12$ & $0.12$ & $0.95$ & $ 0.00$ & $0.12$ & $0.12$ & $0.96$\\
& 12, 24, 36 & $0.03$ & $0.16$ & $0.16$ & $0.96$ & $ 0.01$ & $0.16$ & $0.16$ & $0.97$ & $0.01$ & $0.11$ & $0.11$ & $0.94$ & $ 0.00$ & $0.11$ & $0.10$ & $0.95$\\
\bottomrule
\end{tabular}
\begin{tablenotes}[flushleft]
\item{\em Note:} \textbf{Bias} and \textbf{ESE} are, respectively, the empirical relative bias and standard error of the IUPM estimator; \textbf{ASE} is the average of the standard error estimator; \textbf{CP} is the empirical coverage probability of the 95\% confidence interval for the IUPM. All entries are based on \num{1000} replicates.
\end{tablenotes}
\end{threeparttable}
}
\end{table}

\section{HIV Application}\label{realdata}
\label{s:realdat}

The proposed methods were used to analyze data for 17 people living with HIV on ART with suppressed viral load from the University of North Carolina HIV Cure Center. With multiple dilution QVOA and single dilution UDSA information, these data provide an additional opportunity to quantify the efficiency gain attributable to adding deep sequencing data over QVOA alone. For each subject (i.e., source population), an SLD assay was performed over $D = $ 3--4 dilution levels and with $M = $ 6--36 replicate wells per dilution level. For each subject, deep sequencing was done on 50--100\% of positive wells at one dilution level (i.e., $0.5 \leq q \leq 1$). Details summarizing the assay results are provided in Web Table S4, and the full data are accessible as described in the Data Availability Statement.

Methods applied to the UNC data included the estimators for multiple dilution QVOA with/without UDSA from Section~\ref{methods:multiple_dilutions} and those for single dilution QVOA with UDSA. Previously, \citet{Lee2017} compared the multiple dilution QVOA without UDSA to the single dilution QVOA with UDSA. However, this comparison does not isolate the benefits of using the multiple over single dilution QVOA or the addition of deep sequencing information, since the estimators used either multiple dilutions \textit{or} deep sequencing, but not both. Here, comparisons are made between estimators based on (i) multiple dilution QVOA with versus without UDSA and (ii) single dilution UDSA with single versus multiple dilution QVOA.

\begin{figure}
    \centering
    \includegraphics[width=\textwidth]{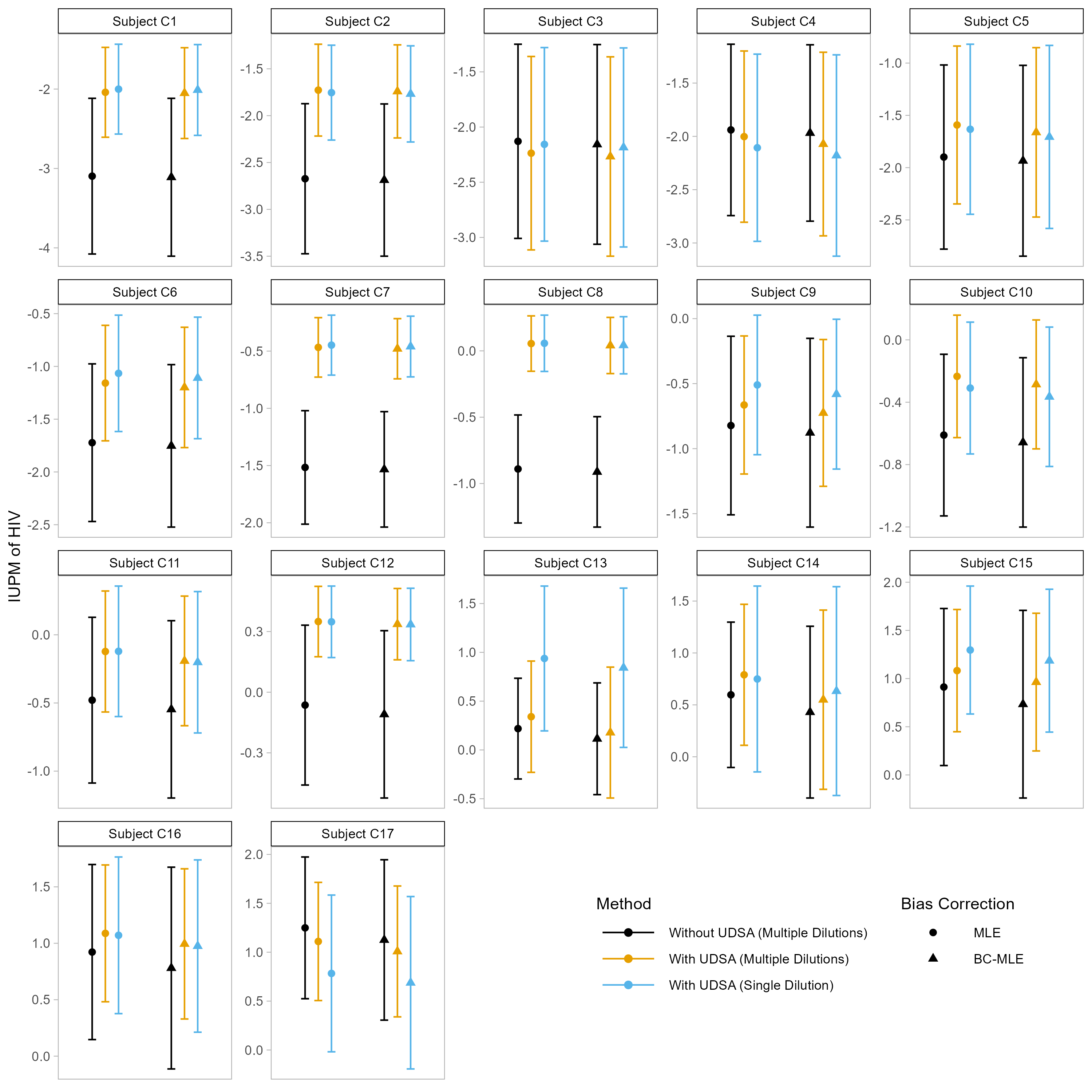}
    \caption{Estimated infectious units per million (IUPM) with 95\% confidence intervals for 17 people living with HIV in the University of North Carolina HIV Cure Center Study. The IUPM and confidence interval were log transformed for comparisons of precision.}
    \label{fig:fig2}
\end{figure}

Estimated log IUPM and 95\% confidence intervals for the 17 people are provided in Figure~\ref{fig:fig2}. The log IUPM and its untransformed confidence interval were used to compare the methods' statistical precision. More detailed analysis results for the IUPM can be found in Web Table S5. As expected, all subjects' bias-corrected IUPM estimates were smaller than their uncorrected ones; these smaller estimates are expected to be closer to the subjects' true HIV concentrations. 

In general, IUPM estimates for a particular subject varied depending on which data were used to compute the estimates. For some subjects (e.g., C7 and C8, who had 39 and 26 DVLs detected), the additional deep sequencing information led to quite different estimates. For other subjects (e.g., C13 and C17), estimates changed substantially when data from a single versus multiple dilutions were used. These differences could perhaps be explained by differences in the subjects' assay results at the unsequenced dilution levels. Subject C17 had a larger estimated IUPM using multiple dilution level data, including a dilution level with all HIV-positive wells, and Subject C13 had a smaller estimated IUPM using multiple dilution level data, including a dilution level with no HIV-positive wells.

Comparing the confidence intervals based on the multiple dilution QVOA with versus without UDSA highlights the precision gain attributable to incorporating the additional sequencing data. For some people, the confidence intervals using the UDSA data were remarkably narrower than those using the QVOA data alone. Incorporating the UDSA data led to the greatest increase in precision for Subject C12, who had 65 observed DVLs across 32 deep sequenced wells and a 57\% narrower confidence interval (for the BC-MLE) when incorporating the UDSA. Meanwhile, for other people with fewer DVLs observed (e.g., Subject C13, who had seven observed DVLs from four sequenced wells), the confidence interval widths did not decrease when using sequencing information. Heuristically, the UDSA is more informative when more DVLs are detected and when more wells are deep sequenced.

Comparing the confidence intervals based on the UDSA with single versus multiple dilution QVOA illustrates the precision gain due to using data from all dilution levels. Again, large gains were seen for some subjects, while the inclusion of multiple dilutions did not change the precision much for others. Consider Subject C17, who had QVOA data available from four dilution levels. For this subject, using data from all dilution levels provided a 25\% narrower confidence interval (for the BC-MLE) than only using data from the one deep sequenced dilution level. On the other hand, consider Subject C1, who had data from three dilution levels and had zero positive wells at the two unsequenced ones. For this person, the two unsequenced, all-negative dilution levels added little information. As a result, the confidence intervals for Subject C1 using UDSA and single or multiple dilution QVOA data are nearly identical. In general, the unsequenced dilutions were more informative when there were many positive wells.

In summary, the estimators based on the multiple dilutions QVOA with UDSA make use of all available information and tend to have better precision than the estimators that ignore the other dilution levels or the UDSA data. The extent of the precision gain depends on the particular assay results.

\section{Overdispersion and Imperfect Assays}
\label{methods:overdispersion-sens-spec}

In the previous sections, it is assumed that (i) the number of cells $X_{ij}$ in well $j$ infected with DVL $i$ follows a Poisson distribution and (ii) that the QVOA and UDSA assays have perfect sensitivity and specificity. This section considers relaxing these two assumptions; details are provided in Web Appendices E and F.

Given that very few cells per well are typically infected with HIV, the Poisson assumption is expected to hold in most settings. Nonetheless, there may be scenarios where the distribution of $X_{ij}$ deviates from a Poisson. In particular, the mean and variance of $X_{ij}$ may not be equal, as the Poisson distribution assumes. To allow for potential overdispersion (i.e., the variance of $X_{ij}$ being greater than its mean), a negative binomial distribution with mean $\lambda_i$ and dispersion parameter $\gamma \in [0, \infty)$ can be used to model $X_{ij}$ wherein $\textrm{Var}(X_{ij}) = \lambda_i + \gamma \lambda_i^2 \geq \lambda_i$. In the special case where $\gamma = 0$, the negative binomial and Poisson distributions are equal.

The observed-data likelihood assuming a negative binomial distribution for $X_{ij}$ is derived in Web Appendix E.1. Given data from multiple dilution levels, this likelihood can be maximized to estimate $\btau$ and $\gamma$. To assess whether there is overdispersion in the assay data, the Poisson and negative binomial MLEs can be used to construct a likelihood ratio test (LRT) of $H_0: \gamma = 0$ vs $H_1: \gamma > 0$. Additional details regarding the overdispersion LRT are provided in Web Appendix E.2. Simulations with overdispersed cell counts demonstrating the performance of the negative binomial MLE and LRT are presented in Web Appendix E.3. Possible overdispersion in the HIV data example from Section~\ref{realdata} is assessed in Web Appendix E.4.

In practice, the assays may have sensitivity or specificity less than 100\%, leading to false positives and negatives. In Web Appendices F.1 and F.2, the single dilution and multiple dilution likelihoods are generalized to allow imperfect sensitivity and specificity of both the QVOA and UDSA. Simulations with imperfect assays are presented in Web Appendix F.3. Across all settings considered, the generalized IUPM estimator that allows for imperfect sensitivity and specificity was empirically unbiased, whereas the IUPM estimator that incorrectly assumed perfect assays was increasingly biased as sensitivity and specificity decreased. Possible imperfect sensitivity and specificity in the HIV data example from Section~\ref{realdata} is assessed in Web Appendix F.4.

\section{Discussion}\label{discuss}
\label{s:discuss}

In this paper, methods were developed to analyze data from SLD assays augmented with additional information provided by deep sequencing. The estimator proposed by \citet{Lee2017}, which uses information from dilution assays and deep sequencing, was given a formal justification and shown to be consistent and asymptotically normal. A bias-corrected MLE was proposed, and it was shown that the MLE is unchanged by the possibility of undetected viral lineages. The \citet{Lee2017} method was extended to the case where the QVOA and deep sequencing data were collected at multiple dilution levels. Additional extensions are provided to relax the Poisson distribution assumption and to accommodate assays with imperfect sensitivity and specificity. Simulations for both the single and multiple dilution settings demonstrated that the BC-MLE has low bias and its corresponding confidence interval achieves nominal coverage. The reduced bias and efficiency gains of the proposed methods relative to existing ones were demonstrated in an application to data from the UNC HIV Cure Center.

There are many directions for future work to expand inference procedures for combined dilution and deep sequencing assays. For the setting with only QVOA data, \citet{Myers1994} derived an exact confidence interval for the IUPM by inverting the likelihood ratio test. \citet{Myers1994} and \citet{Trumble2017} also proposed a goodness-of-fit p-value (PGOF), which can be helpful in identifying possible technical problems with an assay. Calculating both the exact confidence interval and PGOF involve enumerating all possible assay outcomes. Without UDSA data, an assay with $D$ dilution levels and $M^{(d)}$ replicate wells per dilution level $d$ has $\prod_{d=1}^D(M^{(d)}+1)$ possible outcomes. With UDSA data, the number of possible assay outcomes grows much more quickly. For example, if all positive wells were deep sequenced (i.e., $q^{(d)} = 1$ for $d \in \{1, \dots, D\}$), and $n$ DVLs were detected, then there are $\prod_{d=1}^D(M^{(d)}+1)^n$ possible outcomes. This combinatorial explosion makes calculating an exact confidence interval and PGOF for UDSA data computationally challenging. Therefore, developing a computationally feasible exact confidence interval and PGOF to this setting would be interesting areas for future research. Other extensions of the methods considered in this paper include (i) comparing IUPMs between a pair of samples take from an individual before and after a treatment \citep{Li2022} and (ii) comparing the distributions of IUPMs between two treatment groups with multiple individuals per group.

\section*{Acknowledgements}
The authors thank the Associate Editor and an anonymous referee for their helpful comments. This research was supported by the University of North Carolina at Chapel Hill Center for AIDS Research (CFAR), a National Institutes of Health (NIH) funded program P30AI50410, and NIH grant R37AI029168. The content is solely the responsibility of the authors and does not necessarily represent the official views of the NIH. \vspace*{-8pt}


\section*{Supporting Information}

Web Appendices and Tables referenced in Sections 2--5 are available with this paper at the Biometrics website on Wiley Online Library. \texttt{R} code for Sections 3--5 can be found, 
along with the \texttt{R} package \texttt{SLDeepAssay}, 
on GitHub at \url{https://github.com/sarahlotspeich/SLDeepAssay/}. 

\vspace*{-8pt}

\section*{Data Availability Statement}

The data that support the findings of this study are openly available in figshare at \url{https://doi.org/10.6084/m9.figshare.21821229.v1}. 

\vspace*{-8pt}
\bibliographystyle{biom} 
\bibliography{mybib.bib}

\appendix

\section{: Maximum Likelihood Estimation with Undetected Viral Lineages}
\label{s:undected-DVL-proof}

In Section~\ref{methods:undetected}, an augmented function $L'(\blambda|M_N, \bY, Y_0)$ was given that accounts for the $n' - n$ undetected viral lineages. Here, it is proven that the MLE for the mean count of infected cells $\lambda_0$ for the undetected DVLs is necessarily zero.

\begin{proof}
\label{Proof by Contrapositive}

Let $\blambdatilde = (\lambdatilde_0, \dots, \lambdatilde_n)^\T$ be an estimate of $\blambda$ where $\lambdatilde_0 >0$. To show that $\blambdatilde$ cannot be the MLE for $\blambda$, a distinct estimate $\blambdabreve$ will be constructed such that $L'(\blambdabreve|\bY,M_N) > L'(\blambdatilde|\bY,M_N)$. This will prove that any estimate $\blambdatilde$ where $\lambdatilde_0>0$ cannot be the MLE, or, equivalently, that the MLE for $\lambda_0$ \textit{must} be zero.

To construct such an estimate, choose an arbitrary DVL $k$ from those detected by the UDSA, $k \in \left\{1, \dots, n\right\}$. Now, shift some mass $\epsilon \in (0, \lambdatilde_{0})$ from the estimated parameter for the undetected viral lineages to the $k$th parameter, i.e., let  
\begin{align}
\breve{\lambda}_{0} = 
\lambdatilde_{0} - \epsilon \text{ and } \breve{\lambda}_{k} = \lambdatilde_{k} + \epsilon. \label{def_lambda_breve}
\end{align}
For all other DVLs (i.e., $i \in \{1, \dots, n\}$ and $i \neq k$), leave the new estimates unchanged by defining $\lambdabreve_{i} = 
\lambdatilde_{i}$. By construction, $\blambdabreve$ is distinct from $\blambdatilde$ and satisfies $\sum_{i=0}^{n}\lambdabreve_{i} = \sum_{i=0}^{n}\lambdatilde_{i}$.

Now, consider the ratio of the augmented likelihood function \eqref{likelihood_und} evaluated at $\blambdabreve$ and $\blambdatilde$. Many terms cancel out, simplifying this likelihood ratio to 
\begin{align*}
\frac{L\left(\blambdabreve|\bY,M_{N}\right)}{L\left(\blambdatilde|\bY,M_{N}\right)} &= \frac{\left\{1-\exp\left(-\lambdabreve_{k}\right)\right\}^{Y_{k}}\exp\left(-\lambdabreve_{k}\right)^{(M_N+m-Y_{k})}\exp\left(-\lambdabreve_{n+1}\right)^{(M_N+m)}}{\left\{1-\exp\left(-\lambdatilde_{k}\right)\right\}^{Y_{k}}\exp\left(-\lambdahat_{k}\right)^{(M_N+m-Y_{k})}\exp\left(-\lambdatilde_{n+1}\right)^{(M_N+m)}}.
\end{align*}

To show that this likelihood ratio is greater than one, and thus that $\blambdatilde$ cannot be the MLE since it does not maximize the augmented likelihood, recall that $\blambdabreve$ was constructed from $\blambdatilde$. Therefore, using the definitions in \eqref{def_lambda_breve}, the likelihood ratio can be rewritten in terms of only $\blambdatilde$ as 
\begin{align*}
\frac{\left[1-\exp\left\{-\left(\lambdatilde_{k}+\epsilon\right)\right\}\right]^{Y_{k}}\exp\left\{-\left(\lambdatilde_{k}+\epsilon\right)\right\}^{(M_N+m-Y_{k})}\exp\left\{-\left(\lambdatilde_{n+1}-\epsilon\right)\right\}^{(M_N+m)}}{\left\{1-\exp(-\lambdatilde_{k})\right\}^{Y_{k}}\exp(-\lambdatilde_{k})^{(M_N+m-Y_{k})}\exp(-\lambdatilde_{0})^{(M_N+m)}}
\end{align*}
and further simplified to
\begin{align*}
\left\{\frac{\exp\left(\epsilon\right) - \exp\left(-\lambdatilde_{k}\right)}{1 - \exp\left(-\lambdatilde_{k}\right)}\right\}^{Y_k},
\end{align*}
which must be greater than one, completing the proof.
\end{proof}

\end{document}


\maketitle

\section{Finding the MLE}\label{webA}

Note that, for the model assuming perfect assay sensitivity and specificity, the maximum likelihood estimator (MLE) must satisfy $\lambdahat_i \in (0, \infty)$ as long as neither of the extreme assay scenarios from Section 2.3 are met. In this case, an unconstrained optimization method is used to maximize the log-likelihood with respect to $\theta_i = \log(\lambda_i) \in \mathbb{R}$, and the resulting maximizers $\widehat{\theta}_i$ are transformed to obtain the MLEs $\lambdahat_i = \exp(\widehat{\theta}_i) \in (0, \infty)$. In particular, the Broyden, Fletcher, Goldfarb, and Shannon (BFGS) procedure, a quasi-Newton method \citep{bonnans_numerical_2006}, is used to find $\pmb{\widehat{\theta}} = (\widehat{\theta}_1, \dots, \widehat{\theta_n})^T$. This method is implemented via the R \citep{R} function \texttt{optim} using the option \texttt{method="BFGS"}. The initial values are set to $\pmb{\widehat{\theta}}^{(0)} = \log\{-\log\left(1-\bY/M\right)\}$, and the number of iterations is constrained to a maximum of \num{10000}. The gradient of the log-likelihood has the following closed form, 
\begin{align*}
    \nabla l(\blambda|M_N, \bY) &= \begin{bmatrix}
    \frac{\partial}{\partial\lambda_1}l(\blambda|M_N, \bY) & \dots & \frac{\partial}{\partial\lambda_n}l(\blambda|M_N, \bY)
    \end{bmatrix}^\textrm{T}
\end{align*}
with elements given by
\begin{align}
    \frac{\partial}{\partial\lambda_i}l(\blambda|M_N, \bY)
    &= \frac{Y_{i}}{\exp\left(\lambda_{i}\right) - 1}
    -\left(M_{N}+m-Y_{i}\right) + \frac{M-(M_N+m)}{\exp\left(\Lambda\right) - 1}, \label{first_deriv_loglik}
\end{align}
for DVL-specific rate $\lambda_i$, $i \in \{1, \dots, n\}$. Supplying this gradient to the BFGS routine allows for efficient numerical optimization.

The BFGS procedure is guaranteed to converge to a global maximum if the objective function is concave  \citep{bonnans_numerical_2006}. To see that the log-likelihood is in fact concave, note that its second derivatives are
\begin{align}
    \frac{\partial^2}{\partial\lambda_i^2}l(\blambda|M_N,\bY) = & -\frac{Y_{i}\exp\left(\lambda_{i}\right)}{\left\{\exp\left(\lambda_{i}\right) - 1\right\}^2}
    - \frac{\left\{M - \left(M_N+m\right)\right\}\exp\left( \Lambda\right)}{\left\{\exp\left(\Lambda\right) - 1\right\}^{2}}, \label{d2-1} \\
    \frac{\partial^2}{\partial\lambda_i \partial\lambda_j}l(\blambda|M_N,\bY) = & 
    - \frac{\left\{M - \left(M_N+m\right)\right\}\exp\left( \Lambda\right)}{\left\{\exp\left(\Lambda\right) - 1\right\}^{2}}, \label{d2-2}
\end{align}
for $i, j \in \{1, \dots, n\}$. These derivatives can also be written as
\begin{align*}
    \frac{\partial^2}{\partial\blambda^2}l(\blambda|M_N,\bY) = -& \text{diag}\left[\frac{Y_{i}\exp\left(\blambda\right)}{\left\{\exp\left(\blambda\right) - 1\right\}^2}\right] - \frac{\left\{M - \left(M_N+m\right)\right\}\exp\left( \Lambda\right)}{\left\{\exp\left(\Lambda\right) - 1\right\}^{2}} \pmb{J}_n,
\end{align*}
where $\pmb{J}_n$ is an $n \times n$ matrix of ones and $\text{diag}(\pmb{v})$ is an $n \times n$ diagonal matrix with the $n$-vector $\pmb{v}$ on the diagonal. Since $\partial^2l(\blambda|M_N,\bY)/ \partial\blambda^2$ is a sum of a negative definite matrix and a negative semi-definite matrix, it is negative definite. Thus the log-likelihood is concave, guaranteeing global convergence of the BFGS procedure.

As noted in Section 2.3, analytical solutions for $\blambdahat$ do exist in two special cases with data from a single dilution level: (i) when no positive wells are deep sequenced (\ref{sc1}) or (ii) when all positive wells are deep sequenced (\ref{sc2}). 

\subsection{Closed-form MLE when no deep sequencing information is available}\label{sc1}

In case (i), the MLE is equivalent to the IUPM estimator from \citet{Myers1994} and \citet{Trumble2017}, since without the UDSA data (1) in the main text reduces to the likelihood from Myers et al. with $D = 1$ dilution level, i.e., 
\begin{align}
    L(\Lambda|M_N) &= \left\{1 - \exp\left(-\Lambda\right)\right\}^{(M - M_N)}\exp\left(-\Lambda\right)^{M_N}. \label{no_udsa_likelihood}
\end{align}
In fact, in this case the MLE for the IUPM can be found directly, rather than finding $\blambdahat$ first and taking its sum. Taking the derivative of the log-likelihood based on \eqref{no_udsa_likelihood} with respect to $\Lambda$,
\begin{align*}
    \frac{\partial}{\partial\Lambda}l(\Lambda|M_N) &= \frac{M - M_N}{\exp\left(\Lambda\right) - 1} - M_N,
\end{align*}
and finding its root leads to the MLE: $\Lambdahat = \log\left(M/M_N\right)$.

\subsection{Closed-form MLE when deep sequencing information is fully available}\label{sc2}

In case (ii), $m = M_{P}$ and therefore (1) has the simplified form
\begin{align}
L(\blambda|\bY) &= \prod_{i=1}^{n}\left\{1-\exp(-\lambda_i)\right\}^{Y_{i}}\exp(-\lambda_i)^{(M-Y_i)}, 
\end{align}
with corresponding log-likelihood proportional to
\begin{align}
l(\blambda|\bY) &= \sum_{i=1}^{n}Y_{i}\log\left\{1-\exp\left(-\lambda_i\right)\right\} - \sum_{i=1}^{n}(M-Y_i)\lambda_i. \label{loglikelihood1}
\end{align}
The MLE is defined to maximize \eqref{loglikelihood1}; in other words, $\lambdahat_i$ are the roots
of the following derivatives with respect to each DVL-specific rate $\lambda_i$:
\begin{align}
    \frac{\partial}{\partial\lambda_i}l(\blambda|\bY)
    & = Y_i\left\{\frac{\exp\left(-\lambda_{i}\right)}{1-\exp\left(-\lambda_{i}\right)}\right\} - (M-Y_i) \label{loglikelihood1_deriv}
\end{align}
for $i \in \{1, \dots, n\}$. Setting \eqref{loglikelihood1_deriv} equal to zero and solving for $\lambda_i$ yields the MLE: $\widehat{\blambda} = - \log\left(1-\bY/M\right)$.

\section{Derivations for the Variance Estimator}
\label{webB}

In Section 2.3, it was noted that the $n \times n$ covariance matrix $\pmb{\Sigma}$ of the MLE $\blambdahat$ is the inverse of the Fisher Information, which can be estimated with $\widehat{\pmb{\Sigma}} = \sI(\blambda)^{-1}|_{\blambda = \blambdahat}$. In this section, $\sI(\blambda)$ is derived using the log-likelihood based on (1) from the main text. To begin, consider the case where data are available from a single dilution level of one million cells per well, as in Sections 2.1--2.5. In \ref{webD}, an extension to the more general setting with a different single dilution or multiple dilutions is provided.

To compute the Fisher information, expectations of the random quantities $Y_i$, $M_N$, and $m$ are needed. The number of negative wells $M_N$ is a binomial random variable with $M$ trials (i.e., the total number of replicate wells) and probability of success (i.e., of a well being uninfected) equal to $\Pr(W_{j}=0) = \exp\left(-\Lambda\right)$, so $\E\left(M_{N}\right) = M\exp\left(-\Lambda\right)$. Given that $M$ and $q$ are fixed, $m = \nint{q(M - M_N)}$ is a function of $M_N$ and thus
\begin{align*}
    \E(m) &= \sum_{k=0}^M \nint{q(M - k)}\Pr(M_N = k) \\
    &= \sum_{k=0}^M \nint{q(M - k)}{M \choose k} \left\{\exp(-\Lambda)\right\}^k \left\{1 - \exp(-\Lambda)\right\}^{M-k}.
\end{align*}

A similar approach can be used to calculate the expectation of $Y_i$. Recall that for DVL $i$, $i \in \{1, \dots, n\}$, $Y_i = \sum_{j=1}^M X_{ij}R_j$ is defined as the total number of wells that are positive for DVL $i$ (i.e., $X_{ij}=1$) \textit{and} have complete data (i.e., $R_j=1$). Using this definition and the assumption that the DVL indicators $X_{ij}$ in the $M$ wells are independent and identically distributed, the expectation of $Y_i$ is $\E(Y_i) = \sum_{j=1}^M \E(X_{ij}R_j) = M\E(X_{i1}R_1),$ where the first well was chosen here without loss of generality. 

Since $X_{i1}R_1$ is a product of indicator variables, $\E(X_{i1}R_1) = \Pr(X_{i1}R_1=1) = \Pr(X_{i1}=1, R_1=1)$. To compute this probability, it is useful to condition on $M_P$, the number of positive wells, and employ the Law of Total Probability as follows
\begin{align}
    \Pr(X_{i1}=1, R_1=1) = \sum_{k=0}^M \Pr(X_{i1}=1, R_1=1 | M_P=k)\Pr(M_P=k).
\label{EXR}
\end{align}
Under the assumption of perfect sensitivity and specificity for the QVOA (A3), if $X_{i1} = 1$ then $W_{1} = 1$, also. Therefore, $\Pr(X_{i1}=1, R_1=1 | M_P=k) = \Pr(W_1=1, X_{i1}=1, R_1=1 | M_P=k)$, which can be factored as follows 
\begin{align}
    &\Pr(W_1=1, X_{i1}=1, R_1=1 | M_P=k) \nonumber \\
    &= \Pr(R_1=1 | X_{i1} = 1, W_1=1, M_P=k)\Pr(X_{i1}=1| W_1=1, M_P=k)\Pr(W_1=1|M_P=k) \nonumber \\
    &= \left( \frac{\nint{qk}}{k} \right)\left\{\frac{1-\exp(-\lambda_i)}{1-\exp(-\Lambda)} \right\}\left(\frac{k}{M}\right) \nonumber \\
    &= \left(\frac{\nint{qk}}{M}\right)\left\{\frac{1-\exp(-\lambda_i)}{1-\exp(-\Lambda)} \right\}.
    \label{EXR-given-MP}
\end{align}

Combining \eqref{EXR} and \eqref{EXR-given-MP} leads to the expectation of $Y_i$:
\begin{align}
    \E(Y_i) &= M\E(X_{i1}R_1) \nonumber \\
    &= M \sum_{k=0}^M \Pr(M_P=k)\Pr(X_{i1}=1, R_1=1 | M_P=k) \nonumber \\
    &= M \sum_{k=1}^M {M \choose k} \left\{1 - \exp(-\Lambda)\right\}^{k-1} \exp(-\Lambda)^{M-k} \left(\frac{\nint{qk}}{M}\right)\left\{\frac{1-\exp(-\lambda_i)}{1-\exp(-\Lambda)} \right\} \nonumber \\
    &= \left\{1 - \exp(-\lambda_i) \right\} \sum_{k=1}^{M} {M \choose k} \nint{qk} \left\{1 - \exp(-\Lambda)\right\}^{k-1} \exp(-\Lambda)^{M-k}.
    \label{EY}
\end{align}

Since the second derivatives of the log-likelihood are linear in $Y_i$, $M_N$, and $m$, the elements of the Fisher information can by computed by substituting $\E(Y_i)$, $\E(M_N)$, and $\E(m)$ into the negations of \eqref{d2-1} and \eqref{d2-2} to obtain 
\begin{align}
    & \sI_{ii}(\blambda) &= \E\left\{-\frac{\partial^2}{\partial^2\lambda_i^2}l(\blambda|M_N,\bY)\right\} = \frac{\E(Y_i)\exp\left(\lambda_{i}\right)}{\left\{\exp\left(\lambda_{i}\right) - 1\right\}^2}
+ \frac{\left\{M - M\exp(-\Lambda) - \E(m) \right\}\exp\left( \Lambda\right)}{\left\{\exp\left(\Lambda\right) - 1\right\}^{2}}
\label{fisher_ii}
\end{align}
along the diagonal and
\begin{align}
    \sI_{ij}(\blambda) &= \E\left\{-\frac{\partial^2}{\partial\lambda_i \partial\lambda_j}l(\blambda|M_N,\bY)\right\} = \frac{\left\{M - M\exp(-\Lambda) - \E(m) \right\}\exp\left( \Lambda\right)}{\left\{\exp\left(\Lambda\right) - 1\right\}^{2}},
\label{fisher_ij}
\end{align}
on the off-diagonal, respectively, of $\sI(\blambda)$.

\section{Derivations for the Bias Correction}
\label{webC}

In Section~2.4 of the main text, the bias correction for the MLE of the DVL-specific rates was introduced as $B(\blambdahat) = \widehat{\pmb{\Sigma}} ~ \bA(\blambdahat)\mathrm{vec}(\widehat{\pmb{\Sigma}})$, where $\bA\!\left(\blambda\right)  = \left[\bA_1\!\left(\blambda\right) , \dots, \bA_n\!\left(\blambda\right)\right]$ 
 is the $n \times n^2$ matrix with $n \times n$ submatrices
\begin{align*}
\bA_i\!\left(\blambda\right) = \frac{\partial}{\partial \lambda_i}\sI(\blambda) - \frac{1}{2} \textrm{E}\left\{ \frac{\partial^3}{\partial \blambda \partial \blambda ^T \partial \lambda_i}l(\blambda|M_N,\bY) \right\}
\end{align*}
and $\mathrm{vec}\!\left(\pmb{\Sigma}\right)$ denotes the $n^2 \times 1$ column vector obtained by stacking the columns of $\pmb{\Sigma}$. In this section, the necessary components of $\bA_i(\blambda)$ are derived. As with the derivation of the Fisher information in \ref{webB}, it will be useful to first consider the case where data are available from one dilution level of one million cells per well. An extension to the more general setting is provided in \ref{webD}.

Closed form expressions for the Fisher information are given by \eqref{fisher_ii} and \eqref{fisher_ij}, and can be differentiated numerically. Since the expectation of $\partial^3l(\blambda|M_N,\bY)/\partial\lambda_i^3$ is used in $\bA_i(\blambda)$, an analytic expression for the third derivative of the log-likelihood is required. Picking up with the second derivative in \eqref{d2-1} and \eqref{d2-2}, the third derivative has entries
\begin{align*}
    \frac{\partial^3}{\partial\lambda_i^3}l(\blambda|M_N,\bY) &= \frac{Y_{i}\exp\left(\lambda_{i}\right)\left\{\exp\left(\lambda_{i}\right) + 1\right\}}{\left\{\exp\left(\lambda_{i}\right) - 1\right\}^3} + \frac{\left(M - M_N-m\right)\exp\left( \Lambda\right)\left\{\exp\left( \Lambda\right) + 1\right\}}{\left\{\exp\left( \Lambda\right) - 1\right\}^3} \\
    \frac{\partial^3}{\partial\lambda_i^2 \partial\lambda_j}l(\blambda|M_N,\bY) &= \frac{\partial^3}{\partial\lambda_i \partial\lambda_j \partial\lambda_k}l(\blambda|M_N,\bY) = \frac{\left(M - M_N-m\right)\exp\left( \Lambda\right)\left\{\exp\left( \Lambda\right) + 1\right\}}{\left\{\exp\left( \Lambda\right) - 1\right\}^3},
\end{align*}
for distinct $i, j, k \in \{1, \dots, n\}$. Then the expectations of the third derivative are
\begin{align}
    \E \left\{ \frac{\partial^3}{\partial\lambda_i^3}l(\blambda|M_N,\bY)\right\} = & \frac{\E(Y_{i})\exp\left(\lambda_{i}\right)\left\{\exp\left(\lambda_{i}\right) + 1\right\}}{\left\{\exp\left(\lambda_{i}\right) - 1\right\}^3} \nonumber \\
    & + \frac{\left\{M - M\exp(-\Lambda) - \E(m)\right\}\exp\left( \Lambda\right)\left\{\exp\left( \Lambda\right) + 1\right\}}{\left\{\exp\left( \Lambda\right) - 1\right\}^3}, \label{exp-3rd-deriv-1} \\
    \E \left\{ \frac{\partial^3}{\partial\lambda_i^2 \partial\lambda_j}l(\blambda|M_N,\bY)\right\} = & \E \left\{ \frac{\partial^3}{\partial\lambda_i \partial\lambda_j \partial\lambda_k}l(\blambda|M_N,\bY)\right\} \nonumber \\
    = & \frac{\left\{M - M\exp(-\Lambda) - \E(m)\right\}\exp\left( \Lambda\right)\left\{\exp\left( \Lambda\right) + 1\right\}}{\left\{\exp\left( \Lambda\right) - 1\right\}^3},
\label{exp-3rd-deriv-2}
\end{align}
where $\E(m)$ and $\E(Y_i)$ are derived in \ref{webB}. 

\section{Additional Derivations for Multiple Dilution Levels}
\label{webD}

The derivations in Web Appendices B and C can be extended to the multiple dilution level setting as described below. Following the notation introduced in the main text, suppose that assay data $(M_{N}^{(d)}, \bY^{(d)}, u^{(d)})$, $d \in \{1,\dots D\}$, are available from $D$ distinct dilution levels, $D \in \{1, 2, \dots \}$. Based on (4) in the main text, the joint log-likelihood given data from all $D$ dilution levels is proportional to
\begin{align*}
    \widetilde{l}(\btau|\pmb{M_N}, \bY, \pmb{u}) = \sum_{d=1}^{D} l(u^{(d)}\btau|M_{N}^{(d)}, \bY^{(d)}),
\end{align*}
where $l(u\btau|M_{N}, \bY)$, based on (1) in the main text, is proportional to the log-likelihood for a single dilution level. 

Using the chain rule, the first two derivatives 
are
\begin{align*}
    \frac{\partial}{\partial\btau}\widetilde{l}(\btau|\pmb{M_N}, \bY, \pmb{u}) &= \sum_{d=1}^D  u^{(d)} \frac{\partial}{\partial\blambda}l(u^{(d)}\btau|M_{N}^{(d)}, \bY^{(d)}) \\
    \frac{\partial^2}{\partial\btau \partial \btau^\T}\widetilde{l}(\btau|\pmb{M_N}, \bY, \pmb{u}) &= \sum_{d=1}^D \left(u^{(d)}\right)^2 \frac{\partial^2}{\partial\blambda \partial \blambda^\T}l(u^{(d)}\btau|M_{N}^{(d)}, \bY^{(d)}). 
\end{align*}

Using the linearity of expectations, the Fisher information matrix $\widetilde{\sI}(\btau)$ given data from multiple dilutions can be expressed in terms of $\sI(\blambda)$ from \ref{webB}:
\begin{align*}
    \widetilde{\sI}(\btau) = \E\left\{-\frac{\partial^2}{\partial\btau \partial \btau^\T}\widetilde{l}(\btau|\pmb{M_N}, \bY, \pmb{u})\right\} = \sum_{d=1}^D  \left(u^{(d)}\right)^2 \sI(u^{(d)}\btau|M_{N}^{(d)}, \bY^{(d)}),
\end{align*}
where $\sI(u^{(d)}\btau|M_{N}^{(d)}, \bY^{(d)})$ denotes the previously developed Fisher information matrix for $\blambda$ evaluated at $(\blambda = u^{(d)}\btau, M_N = M_{N}^{(d)}, \bY = \bY^{(d)})$. Then the covariance matrix $\widetilde{\pmb{\Sigma}}$ of $\btau$ can be estimated with $\widetilde{\sI}^{-1}(\btau)|_{\btau = \btauhat}$.

A bias correction term for the MLE $\btauhat$ given data from multiple dilution levels is
\begin{align}
    \widetilde{B}\left(\btauhat\right) = \left\{\widetilde{\sI}^{-1}(\btau)|_{\btau = \btauhat}\right\}
    \widetilde{\bA} \left(\btauhat\right) \mathrm{vec}\left\{\widetilde{\sI}^{-1}(\btau)|_{\btau = \btauhat}\right\},
\end{align}
where $\widetilde{\bA}(\btau)  = [\widetilde{\bA}_1(\btau) , \dots, \widetilde{\bA}_n(\btau)]$ is the $n \times n^2$ matrix with $n \times n$ submatrices defined as
\begin{align*}
    \widetilde{\bA}_i\left(\btau\right) = \frac{\partial}{\partial\tau_i}\widetilde{\sI}(\btau) - \frac{1}{2} \textrm{E}\left\{ \frac{\partial^3}{\partial \btau \partial \btau ^\T \partial \tau_i}\widetilde{l}(\btau|\pmb{M_N},\bY, \pmb{u})\right\}.
\end{align*}

As in the single dilution case, the derivative $\partial \widetilde{\sI}(\btau)/\partial \tau_i$ can be computed numerically. Also, using the linearity of expectations,
\begin{align*}
    {\rm E}\left\{ \frac{\partial^3 }{\partial \btau \partial \btau ^T \partial \tau_i}\tilde{l}(\btauhat|\pmb{M_N}, \bY, \pmb{u}) \right\} = \sum_{d=1}^D  \left(u^{(d)}\right)^3 {\rm E} \left\{ \frac{\partial^3}{\partial\blambda \partial \blambda^T \partial \lambda_i}l(u^{(d)}\btauhat|M_{N}^{(d)}, \bY^{(d)}) \right\},
\end{align*}
where the expectations inside the sum on the right-hand side can be computed using \eqref{exp-3rd-deriv-1} and \eqref{exp-3rd-deriv-2} from \ref{webC}.

\begin{table}
\centering
\caption{Simulation results with a single dilution level and a non-constant rate of infected cells for all distinct viral lineages. The true overall IUPM in all settings was $T = 1$.\label{TableS1}}
\resizebox{\columnwidth}{!}{
\begin{threeparttable}
\begin{tabular}{cccrrrrrrrrrrrrrrrr}
\toprule
\multicolumn{3}{c}{\textbf{ }} & \multicolumn{8}{c}{\textbf{Without UDSA}} & \multicolumn{8}{c}{\textbf{With UDSA}} \\
\cmidrule(l{3pt}r{3pt}){4-11} \cmidrule(l{3pt}r{3pt}){12-19}
\multicolumn{3}{c}{\textbf{ }} & \multicolumn{4}{c}{\textbf{MLE}} & \multicolumn{4}{c}{\textbf{Bias-Corrected MLE}} & \multicolumn{4}{c}{\textbf{MLE}} & \multicolumn{4}{c}{\textbf{Bias-Corrected MLE}} \\
\cmidrule(l{3pt}r{3pt}){4-7} \cmidrule(l{3pt}r{3pt}){8-11} \cmidrule(l{3pt}r{3pt}){12-15} \cmidrule(l{3pt}r{3pt}){16-19}
\textbf{$\pmb{n'}$} & $\pmb{M}$ & $\pmb{q}$ & \textbf{Bias} & \textbf{ASE} & \textbf{ESE} & \textbf{CP} & \textbf{Bias} & \textbf{ASE} & \textbf{ESE} & \textbf{CP} & \textbf{Bias} & \textbf{ASE} & \textbf{ESE} & \textbf{CP} & \textbf{Bias} & \textbf{ASE} & \textbf{ESE} & \textbf{CP}\\
\midrule
6 & 12 & 0.50 & $0.10$ & $0.42$ & $0.45$ & $0.95$ & $ 0.00$ & $0.42$ & $0.37$ & $0.90$ & $0.09$ & $0.36$ & $0.38$ & $0.94$ & $-0.02$ & $0.36$ & $0.34$ & $0.98$\\
& & 0.75 & $0.10$ & $0.42$ & $0.45$ & $0.95$ & $ 0.00$ & $0.42$ & $0.37$ & $0.90$ & $0.07$ & $0.34$ & $0.35$ & $0.94$ & $ 0.00$ & $0.34$ & $0.33$ & $0.97$\\
& & 1.00 & $0.10$ & $0.42$ & $0.45$ & $0.95$ & $ 0.00$ & $0.42$ & $0.37$ & $0.90$ & $0.05$ & $0.32$ & $0.33$ & $0.95$ & $ 0.00$ & $0.32$ & $0.31$ & $0.97$\\
\addlinespace
& 24 & 0.50 & $0.05$ & $0.28$ & $0.30$ & $0.96$ & $ 0.01$ & $0.28$ & $0.28$ & $0.94$ & $0.04$ & $0.25$ & $0.25$ & $0.95$ & $-0.01$ & $0.25$ & $0.24$ & $0.97$\\
& & 0.75 & $0.05$ & $0.28$ & $0.30$ & $0.96$ & $ 0.01$ & $0.28$ & $0.28$ & $0.94$ & $0.03$ & $0.23$ & $0.24$ & $0.94$ & $ 0.00$ & $0.23$ & $0.23$ & $0.95$\\
& & 1.00 & $0.05$ & $0.28$ & $0.30$ & $0.96$ & $ 0.01$ & $0.28$ & $0.28$ & $0.94$ & $0.02$ & $0.22$ & $0.23$ & $0.95$ & $ 0.00$ & $0.22$ & $0.22$ & $0.96$\\
\addlinespace
& 32 & 0.50 & $0.03$ & $0.24$ & $0.25$ & $0.96$ & $ 0.00$ & $0.24$ & $0.24$ & $0.91$ & $0.03$ & $0.21$ & $0.22$ & $0.94$ & $ 0.00$ & $0.21$ & $0.21$ & $0.96$\\
& & 0.75 & $0.03$ & $0.24$ & $0.25$ & $0.96$ & $ 0.00$ & $0.24$ & $0.24$ & $0.91$ & $0.02$ & $0.20$ & $0.20$ & $0.94$ & $ 0.00$ & $0.20$ & $0.20$ & $0.96$\\
& & 1.00 & $0.03$ & $0.24$ & $0.25$ & $0.96$ & $ 0.00$ & $0.24$ & $0.24$ & $0.91$ & $0.02$ & $0.19$ & $0.19$ & $0.94$ & $ 0.00$ & $0.19$ & $0.19$ & $0.95$\\
\addlinespace
12 & 12 & 0.50 & $0.09$ & $0.41$ & $0.45$ & $0.96$ & $-0.01$ & $0.41$ & $0.37$ & $0.89$ & $0.08$ & $0.35$ & $0.37$ & $0.94$ & $-0.02$ & $0.35$ & $0.34$ & $0.98$\\
& & 0.75 & $0.09$ & $0.41$ & $0.45$ & $0.96$ & $-0.01$ & $0.41$ & $0.37$ & $0.89$ & $0.06$ & $0.33$ & $0.34$ & $0.94$ & $-0.01$ & $0.33$ & $0.32$ & $0.96$\\
& & 1.00 & $0.09$ & $0.41$ & $0.45$ & $0.96$ & $-0.01$ & $0.41$ & $0.37$ & $0.89$ & $0.04$ & $0.31$ & $0.31$ & $0.95$ & $-0.01$ & $0.31$ & $0.29$ & $0.97$\\
\addlinespace
& 24 & 0.50 & $0.04$ & $0.28$ & $0.31$ & $0.96$ & $ 0.00$ & $0.28$ & $0.28$ & $0.93$ & $0.03$ & $0.24$ & $0.24$ & $0.94$ & $-0.01$ & $0.24$ & $0.23$ & $0.96$\\
& & 0.75 & $0.04$ & $0.28$ & $0.31$ & $0.96$ & $ 0.00$ & $0.28$ & $0.28$ & $0.93$ & $0.03$ & $0.23$ & $0.23$ & $0.95$ & $ 0.00$ & $0.23$ & $0.22$ & $0.96$\\
& & 1.00 & $0.04$ & $0.28$ & $0.31$ & $0.96$ & $ 0.00$ & $0.28$ & $0.28$ & $0.93$ & $0.02$ & $0.21$ & $0.21$ & $0.95$ & $-0.01$ & $0.21$ & $0.21$ & $0.96$\\
\addlinespace
& 32 & 0.50 & $0.02$ & $0.24$ & $0.24$ & $0.95$ & $-0.01$ & $0.24$ & $0.23$ & $0.92$ & $0.02$ & $0.21$ & $0.21$ & $0.95$ & $-0.01$ & $0.21$ & $0.20$ & $0.95$\\
& & 0.75 & $0.02$ & $0.24$ & $0.24$ & $0.95$ & $-0.01$ & $0.24$ & $0.23$ & $0.92$ & $0.02$ & $0.19$ & $0.20$ & $0.94$ & $-0.01$ & $0.19$ & $0.19$ & $0.95$\\
& & 1.00 & $0.02$ & $0.24$ & $0.24$ & $0.95$ & $-0.01$ & $0.24$ & $0.23$ & $0.92$ & $0.01$ & $0.18$ & $0.18$ & $0.95$ & $ 0.00$ & $0.18$ & $0.18$ & $0.95$\\
\addlinespace
18 & 12 & 0.50 & $0.06$ & $0.41$ & $0.42$ & $0.97$ & $-0.03$ & $0.41$ & $0.35$ & $0.89$ & $0.07$ & $0.35$ & $0.35$ & $0.95$ & $-0.02$ & $0.35$ & $0.32$ & $0.98$\\
& & 0.75 & $0.06$ & $0.41$ & $0.42$ & $0.97$ & $-0.03$ & $0.41$ & $0.35$ & $0.89$ & $0.05$ & $0.32$ & $0.32$ & $0.96$ & $-0.02$ & $0.32$ & $0.30$ & $0.97$\\
& & 1.00 & $0.06$ & $0.41$ & $0.42$ & $0.97$ & $-0.03$ & $0.41$ & $0.35$ & $0.89$ & $0.03$ & $0.30$ & $0.30$ & $0.95$ & $-0.01$ & $0.30$ & $0.28$ & $0.98$\\
\addlinespace
& 24 & 0.50 & $0.03$ & $0.28$ & $0.30$ & $0.96$ & $-0.01$ & $0.28$ & $0.28$ & $0.93$ & $0.03$ & $0.24$ & $0.25$ & $0.94$ & $-0.01$ & $0.24$ & $0.24$ & $0.96$\\
& & 0.75 & $0.03$ & $0.28$ & $0.30$ & $0.96$ & $-0.01$ & $0.28$ & $0.28$ & $0.93$ & $0.02$ & $0.22$ & $0.23$ & $0.94$ & $-0.01$ & $0.22$ & $0.22$ & $0.96$\\
& & 1.00 & $0.03$ & $0.28$ & $0.30$ & $0.96$ & $-0.01$ & $0.28$ & $0.28$ & $0.93$ & $0.02$ & $0.21$ & $0.22$ & $0.94$ & $-0.01$ & $0.21$ & $0.21$ & $0.95$\\
\addlinespace
& 32 & 0.50 & $0.03$ & $0.24$ & $0.25$ & $0.96$ & $ 0.00$ & $0.24$ & $0.23$ & $0.90$ & $0.03$ & $0.21$ & $0.21$ & $0.94$ & $ 0.00$ & $0.21$ & $0.20$ & $0.96$\\
& & 0.75 & $0.03$ & $0.24$ & $0.25$ & $0.96$ & $ 0.00$ & $0.24$ & $0.23$ & $0.90$ & $0.02$ & $0.19$ & $0.20$ & $0.94$ & $ 0.00$ & $0.19$ & $0.19$ & $0.95$\\
& & 1.00 & $0.03$ & $0.24$ & $0.25$ & $0.96$ & $ 0.00$ & $0.24$ & $0.23$ & $0.90$ & $0.02$ & $0.18$ & $0.19$ & $0.94$ & $ 0.00$ & $0.18$ & $0.19$ & $0.95$\\
\bottomrule
\end{tabular}
\begin{tablenotes}[flushleft]
\item{\em Note:} \textbf{Bias} and \textbf{ESE} are, respectively, the empirical relative bias and standard error of the IUPM estimator; \textbf{ASE} is the average of the standard error estimator; \textbf{CP} is the empirical coverage probability of the 95\% confidence interval for the IUPM. There were a total of 39 simulated assays out of \num{27000} ($0.1\%$) where the MLE and bias-corrected MLE without UDSA were infinite that were excluded; all other entries are based on \num{1000} replicates.
\end{tablenotes}
\end{threeparttable}
}
\end{table}

\begin{table}
\centering
\caption{Simulation results with a single dilution level and a constant rate of infected cells for all distinct viral lineages. The true overall IUPM in all settings was $T = 0.5$.\label{TableS2}}
\resizebox{\columnwidth}{!}{
\begin{threeparttable}
\begin{tabular}{cccrrrrrrrrrrrrrrrr}
\toprule
\multicolumn{3}{c}{\textbf{ }} & \multicolumn{8}{c}{\textbf{Without UDSA}} & \multicolumn{8}{c}{\textbf{With UDSA}} \\
\cmidrule(l{3pt}r{3pt}){4-11} \cmidrule(l{3pt}r{3pt}){12-19}
\multicolumn{3}{c}{\textbf{ }} & \multicolumn{4}{c}{\textbf{MLE}} & \multicolumn{4}{c}{\textbf{Bias-Corrected MLE}} & \multicolumn{4}{c}{\textbf{MLE}} & \multicolumn{4}{c}{\textbf{Bias-Corrected MLE}} \\
\cmidrule(l{3pt}r{3pt}){4-7} \cmidrule(l{3pt}r{3pt}){8-11} \cmidrule(l{3pt}r{3pt}){12-15} \cmidrule(l{3pt}r{3pt}){16-19}
\textbf{$\pmb{n'}$} & $\pmb{M}$ & $\pmb{q}$ & \textbf{Bias} & \textbf{ASE} & \textbf{ESE} & \textbf{CP} & \textbf{Bias} & \textbf{ASE} & \textbf{ESE} & \textbf{CP} & \textbf{Bias} & \textbf{ASE} & \textbf{ESE} & \textbf{CP} & \textbf{Bias} & \textbf{ASE} & \textbf{ESE} & \textbf{CP}\\
\midrule
6 & 12 & 0.50 & $0.03$ & $0.24$ & $0.26$ & $0.95$ & $ 0.00$ & $0.24$ & $0.23$ & $0.75$ & $0.04$ & $0.23$ & $0.24$ & $0.95$ & $-0.02$ & $0.23$ & $0.22$ & $0.98$\\
& & 0.75 & $0.03$ & $0.24$ & $0.26$ & $0.95$ & $ 0.00$ & $0.24$ & $0.23$ & $0.75$ & $0.03$ & $0.22$ & $0.23$ & $0.96$ & $-0.01$ & $0.22$ & $0.22$ & $0.98$\\
& & 1.00 & $0.03$ & $0.24$ & $0.26$ & $0.95$ & $ 0.00$ & $0.24$ & $0.23$ & $0.75$ & $0.02$ & $0.21$ & $0.22$ & $0.95$ & $ 0.00$ & $0.21$ & $0.21$ & $0.98$\\
\addlinespace
& 24 & 0.50 & $0.01$ & $0.17$ & $0.17$ & $0.94$ & $ 0.00$ & $0.17$ & $0.17$ & $0.76$ & $0.01$ & $0.16$ & $0.16$ & $0.95$ & $-0.01$ & $0.16$ & $0.16$ & $0.97$\\
& & 0.75 & $0.01$ & $0.17$ & $0.17$ & $0.94$ & $ 0.00$ & $0.17$ & $0.17$ & $0.76$ & $0.01$ & $0.15$ & $0.16$ & $0.95$ & $-0.01$ & $0.15$ & $0.15$ & $0.97$\\
& & 1.00 & $0.01$ & $0.17$ & $0.17$ & $0.94$ & $ 0.00$ & $0.17$ & $0.17$ & $0.76$ & $0.01$ & $0.15$ & $0.15$ & $0.95$ & $ 0.00$ & $0.15$ & $0.15$ & $0.96$\\
\addlinespace
& 32 & 0.50 & $0.01$ & $0.14$ & $0.15$ & $0.95$ & $ 0.00$ & $0.14$ & $0.14$ & $0.69$ & $0.01$ & $0.14$ & $0.14$ & $0.94$ & $-0.01$ & $0.14$ & $0.13$ & $0.95$\\
& & 0.75 & $0.01$ & $0.14$ & $0.15$ & $0.95$ & $ 0.00$ & $0.14$ & $0.14$ & $0.69$ & $0.01$ & $0.13$ & $0.13$ & $0.94$ & $ 0.00$ & $0.13$ & $0.13$ & $0.95$\\
& & 1.00 & $0.01$ & $0.14$ & $0.15$ & $0.95$ & $ 0.00$ & $0.14$ & $0.14$ & $0.69$ & $0.01$ & $0.13$ & $0.13$ & $0.94$ & $ 0.00$ & $0.13$ & $0.13$ & $0.95$\\
\addlinespace
12 & 12 & 0.50 & $0.02$ & $0.24$ & $0.25$ & $0.95$ & $-0.01$ & $0.24$ & $0.23$ & $0.76$ & $0.03$ & $0.23$ & $0.23$ & $0.95$ & $-0.03$ & $0.23$ & $0.21$ & $0.98$\\
& & 0.75 & $0.02$ & $0.24$ & $0.25$ & $0.95$ & $-0.01$ & $0.24$ & $0.23$ & $0.76$ & $0.02$ & $0.22$ & $0.22$ & $0.96$ & $-0.01$ & $0.22$ & $0.21$ & $0.98$\\
& & 1.00 & $0.02$ & $0.24$ & $0.25$ & $0.95$ & $-0.01$ & $0.24$ & $0.23$ & $0.76$ & $0.02$ & $0.21$ & $0.21$ & $0.97$ & $ 0.00$ & $0.21$ & $0.20$ & $0.97$\\
\addlinespace
& 24 & 0.50 & $0.01$ & $0.17$ & $0.17$ & $0.93$ & $ 0.00$ & $0.17$ & $0.17$ & $0.73$ & $0.02$ & $0.16$ & $0.16$ & $0.95$ & $-0.01$ & $0.16$ & $0.15$ & $0.96$\\
& & 0.75 & $0.01$ & $0.17$ & $0.17$ & $0.93$ & $ 0.00$ & $0.17$ & $0.17$ & $0.73$ & $0.01$ & $0.15$ & $0.15$ & $0.96$ & $-0.01$ & $0.15$ & $0.15$ & $0.97$\\
& & 1.00 & $0.01$ & $0.17$ & $0.17$ & $0.93$ & $ 0.00$ & $0.17$ & $0.17$ & $0.73$ & $0.01$ & $0.15$ & $0.15$ & $0.96$ & $ 0.00$ & $0.15$ & $0.14$ & $0.97$\\
\addlinespace
& 32 & 0.50 & $0.01$ & $0.14$ & $0.15$ & $0.95$ & $ 0.00$ & $0.14$ & $0.14$ & $0.70$ & $0.01$ & $0.14$ & $0.14$ & $0.95$ & $-0.01$ & $0.14$ & $0.13$ & $0.96$\\
& & 0.75 & $0.01$ & $0.14$ & $0.15$ & $0.95$ & $ 0.00$ & $0.14$ & $0.14$ & $0.70$ & $0.01$ & $0.13$ & $0.13$ & $0.95$ & $ 0.00$ & $0.13$ & $0.13$ & $0.96$\\
& & 1.00 & $0.01$ & $0.14$ & $0.15$ & $0.95$ & $ 0.00$ & $0.14$ & $0.14$ & $0.70$ & $0.01$ & $0.13$ & $0.13$ & $0.96$ & $ 0.00$ & $0.13$ & $0.12$ & $0.96$\\
\addlinespace
18 & 12 & 0.50 & $0.02$ & $0.24$ & $0.25$ & $0.95$ & $-0.01$ & $0.24$ & $0.23$ & $0.76$ & $0.03$ & $0.23$ & $0.23$ & $0.95$ & $-0.02$ & $0.23$ & $0.22$ & $0.97$\\
& & 0.75 & $0.02$ & $0.24$ & $0.25$ & $0.95$ & $-0.01$ & $0.24$ & $0.23$ & $0.76$ & $0.02$ & $0.22$ & $0.22$ & $0.96$ & $-0.01$ & $0.22$ & $0.21$ & $0.97$\\
& & 1.00 & $0.02$ & $0.24$ & $0.25$ & $0.95$ & $-0.01$ & $0.24$ & $0.23$ & $0.76$ & $0.02$ & $0.21$ & $0.21$ & $0.96$ & $-0.01$ & $0.21$ & $0.20$ & $0.97$\\
\addlinespace
& 24 & 0.50 & $0.01$ & $0.17$ & $0.17$ & $0.94$ & $-0.01$ & $0.17$ & $0.17$ & $0.73$ & $0.01$ & $0.16$ & $0.16$ & $0.95$ & $-0.01$ & $0.16$ & $0.16$ & $0.96$\\
& & 0.75 & $0.01$ & $0.17$ & $0.17$ & $0.94$ & $-0.01$ & $0.17$ & $0.17$ & $0.73$ & $0.01$ & $0.15$ & $0.15$ & $0.95$ & $-0.01$ & $0.15$ & $0.15$ & $0.96$\\
& & 1.00 & $0.01$ & $0.17$ & $0.17$ & $0.94$ & $-0.01$ & $0.17$ & $0.17$ & $0.73$ & $0.01$ & $0.15$ & $0.15$ & $0.96$ & $ 0.00$ & $0.15$ & $0.15$ & $0.96$\\
\addlinespace
& 32 & 0.50 & $0.01$ & $0.14$ & $0.15$ & $0.94$ & $ 0.00$ & $0.14$ & $0.14$ & $0.70$ & $0.01$ & $0.14$ & $0.13$ & $0.94$ & $ 0.00$ & $0.14$ & $0.13$ & $0.95$\\
& & 0.75 & $0.01$ & $0.14$ & $0.15$ & $0.94$ & $ 0.00$ & $0.14$ & $0.14$ & $0.70$ & $0.01$ & $0.13$ & $0.13$ & $0.94$ & $ 0.00$ & $0.13$ & $0.13$ & $0.96$\\
& & 1.00 & $0.01$ & $0.14$ & $0.15$ & $0.94$ & $ 0.00$ & $0.14$ & $0.14$ & $0.70$ & $0.01$ & $0.13$ & $0.13$ & $0.96$ & $ 0.00$ & $0.13$ & $0.13$ & $0.96$\\
\bottomrule
\end{tabular}
\begin{tablenotes}[flushleft]
\item{\em Note:} \textbf{Bias} and \textbf{ESE} are, respectively, the empirical relative bias and standard error of the IUPM estimator; \textbf{ASE} is the average of the standard error estimator; \textbf{CP} is the empirical coverage probability of the 95\% confidence interval for the IUPM. All entries are based on \num{1000} replicates.
\end{tablenotes}
\end{threeparttable}
}
\end{table}

\begin{table}
\centering
\caption{Simulation results with multiple dilution levels and a non-constant rate of infected cells for all distinct viral lineages. In all settings, the true IUPM was $T = 1$ and the proportions of p24-positive wells that were deep-sequenced at the three dilution levels were $\pmb{q} = (0, 0.5, 1)$.\label{TableS3}}
\resizebox{\columnwidth}{!}{
\begin{threeparttable}
\begin{tabular}{ccccrrrrrrrrrrrrrr}
\toprule
\multicolumn{2}{c}{\textbf{ }} & \multicolumn{8}{c}{\textbf{Without UDSA}} & \multicolumn{8}{c}{\textbf{With UDSA}} \\
\cmidrule(l{3pt}r{3pt}){3-10} \cmidrule(l{3pt}r{3pt}){11-18}
\multicolumn{2}{c}{\textbf{ }} & \multicolumn{4}{c}{\textbf{MLE}} & \multicolumn{4}{c}{\textbf{Bias-Corrected MLE}} & \multicolumn{4}{c}{\textbf{MLE}} & \multicolumn{4}{c}{\textbf{Bias-Corrected MLE}} \\
\cmidrule(l{3pt}r{3pt}){3-6} \cmidrule(l{3pt}r{3pt}){7-10} \cmidrule(l{3pt}r{3pt}){11-14} \cmidrule(l{3pt}r{3pt}){15-18}
$\pmb{n'}$ & $\pmb{M}$ & \textbf{Bias} & \textbf{ASE} & \textbf{ESE} & \textbf{CP} & \textbf{Bias} & \textbf{ASE} & \textbf{ESE} & \textbf{CP} & \textbf{Bias} & \textbf{ASE} & \textbf{ESE} & \textbf{CP} & \textbf{Bias} & \textbf{ASE} & \textbf{ESE} & \textbf{CP}\\
\midrule
6 & 6, 12, 18 & $0.04$ & $0.22$ & $0.24$ & $0.95$ & $ 0.00$ & $0.22$ & $0.22$ & $0.96$ & $0.02$ & $0.16$ & $0.16$ & $0.94$ & $-0.01$ & $0.16$ & $0.16$ & $0.95$\\
& 9, 18, 27 & $0.03$ & $0.18$ & $0.19$ & $0.95$ & $ 0.01$ & $0.18$ & $0.18$ & $0.96$ & $0.02$ & $0.13$ & $0.13$ & $0.96$ & $ 0.00$ & $0.13$ & $0.13$ & $0.96$\\
& 12, 24, 36 & $0.02$ & $0.16$ & $0.16$ & $0.95$ & $ 0.00$ & $0.16$ & $0.16$ & $0.96$ & $0.01$ & $0.11$ & $0.11$ & $0.96$ & $ 0.00$ & $0.11$ & $0.11$ & $0.96$\\
\addlinespace
12 & 6, 12, 18 & $0.03$ & $0.22$ & $0.24$ & $0.96$ & $-0.01$ & $0.22$ & $0.22$ & $0.96$ & $0.02$ & $0.15$ & $0.16$ & $0.95$ & $ 0.00$ & $0.15$ & $0.15$ & $0.95$\\
& 9, 18, 27 & $0.03$ & $0.18$ & $0.18$ & $0.96$ & $ 0.00$ & $0.18$ & $0.18$ & $0.96$ & $0.02$ & $0.13$ & $0.13$ & $0.94$ & $ 0.00$ & $0.13$ & $0.12$ & $0.95$\\
& 12, 24, 36 & $0.02$ & $0.16$ & $0.16$ & $0.95$ & $ 0.00$ & $0.16$ & $0.16$ & $0.96$ & $0.01$ & $0.11$ & $0.11$ & $0.95$ & $ 0.00$ & $0.11$ & $0.11$ & $0.95$\\
\addlinespace
18 & 9, 18, 27 & $0.02$ & $0.18$ & $0.19$ & $0.95$ & $ 0.00$ & $0.18$ & $0.18$ & $0.95$ & $0.01$ & $0.12$ & $0.12$ & $0.95$ & $ 0.00$ & $0.12$ & $0.12$ & $0.95$\\
& 6, 12, 18 & $0.04$ & $0.22$ & $0.24$ & $0.96$ & $ 0.00$ & $0.22$ & $0.22$ & $0.96$ & $0.02$ & $0.15$ & $0.16$ & $0.94$ & $ 0.00$ & $0.15$ & $0.15$ & $0.95$\\
& 12, 24, 36  & $0.02$ & $0.16$ & $0.16$ & $0.96$ & $ 0.00$ & $0.16$ & $0.15$ & $0.96$ & $0.01$ & $0.11$ & $0.11$ & $0.94$ & $ 0.00$ & $0.11$ & $0.11$ & $0.95$\\
\bottomrule
\end{tabular}
\begin{tablenotes}[flushleft]
\item{\em Note:} \textbf{Bias} and \textbf{ESE} are, respectively, the empirical relative bias and standard error of the IUPM estimator; \textbf{ASE} is the average of the standard error estimator; \textbf{CP} is the empirical coverage probability of the 95\% confidence interval for the IUPM. All entries are based on \num{1000} replicates.
\end{tablenotes}
\end{threeparttable}
}
\end{table}

\begin{table}
\centering
\caption{Summary of assay data collected from 17 individuals receiving care at the University of North Carolina HIV Cure Center\label{TableS4}}
\fontsize{10}{12}\selectfont
\begin{threeparttable}
\begin{tabular}{ccccccc}
\toprule
\multicolumn{2}{c}{\textbf{ }} & \multicolumn{2}{c}{\textbf{QVOA}} & \multicolumn{2}{c}{\textbf{UDSA}} \\
\cmidrule(l{3pt}r{3pt}){3-4} \cmidrule(l{3pt}r{3pt}){5-6} 
\textbf{ID} & $\pmb{u}$ & $\pmb{M}$ & $\pmb{M_P}$ & $\pmb{m}$ & $\pmb{n}$ \\
\midrule
C1 & $(2.5, 0.5, 0.1)$ & $(36, 6, 6)$ & $(4, 0, 0)$ & $(4, 0, 0)$ & 12\\
C2 & $(2.5, 0.5, 0.1)$ & $(36, 6, 6)$ & $(5, 1, 0)$ & $(5, 0, 0)$ & 7\\
C3 & $(2.5, 0.5, 0.1)$ & $(18, 6, 6)$ & $(5, 0, 0)$ & $(5, 0, 0)$ & 4\\
C4 & $(2.5, 0.5, 0.1)$ & $(18, 6, 6)$ & $(5, 1, 0)$ & $(3, 0, 0)$ & 3\\
C5 & $(2.5, 0.5, 0.1, 0.025)$ & $(14, 6, 6, 6)$ & $(4, 0, 1, 0)$ & $(3, 0, 0, 0)$ & 3\\
C6 & $(2.5, 0.5, 0.1)$ & $(18, 6, 6)$ & $(7, 0, 0)$ & $(6, 0, 0)$ & 4\\
C7 & $(2.5, 0.5, 0.1)$ & $(36, 6, 6)$ & $(15, 1, 0)$ & $(15, 0, 0)$ & 39 \\
C8 & $(2.5, 0.5, 0.1)$ & $(36, 6, 6)$ & $(22, 2, 1)$ & $(22, 0, 0)$ & 26 \\
C9 & $(2.5, 0.5, 0.1, 0.025)$ & $(12, 6, 6, 6)$ & $(9, 0, 0, 0)$ & $(6, 0, 0, 0)$ & 8 \\
C10 & $(2.5, 0.5, 0.1)$ &$(18, 6, 6)$ & $(12, 3, 1)$ & $(6, 0, 0)$ & 19 \\
C11 & $(2.5, 0.5, 0.1, 0.025)$ & $(12, 6, 6, 6)$ & $(9, 1, 1, 1)$ & $(6, 0, 0, 0)$ & 9 \\
C12 & $(2.5, 0.5, 0.1)$ & $(36, 6, 6)$ & $(32, 3, 1)$ & $(32, 0, 0)$ & 65 \\
C13 & $(2.5, 0.5, 0.1, 0.025)$ & $(18, 6, 6, 6)$ & $(16, 4, 3, 0)$ & $(0, 4, 0, 0)$ & 7 \\
C14 & $(2.5, 0.5, 0.1)$ & $(18, 6, 6)$ & $(18, 3, 1)$ & $(0, 3, 0)$ & 3\\
C15 & $(2.5, 0.5, 0.1)$ & $(18, 6, 6)$ & $(18, 5, 0)$ & $(0, 5, 0)$ & 6\\
C16 & $(2.5, 0.5, 0.1, 0.025)$ & $(12, 6, 6, 6)$ & $(12, 4, 2, 0)$ & $(0, 4, 0, 0)$ & 8\\
C17 & $(2.5, 0.5, 0.1, 0.025)$ & $(18, 6, 6, 6)$ & $(18, 4, 3, 1)$ & $(0, 4, 0, 0)$ & 6\\
\bottomrule
\end{tabular}
\begin{tablenotes}[flushleft]
\footnotesize{\item{\em Note:} $\pmb{u}$ is the vector of dilution levels in millions of cells per well; $\pmb{M}$ is the vector of replicate well numbers to undergo the QVOA at each dilution level; $\pmb{M_P}$ is the vector of positive wells found by the QVOA at each dilution level; $\pmb{m}$ is the vector of positive well numbers to undergo the UDSA at each dilution level; $\pmb{n}$ is the number of distinct viral lineages (DVL) detected by the UDSA.} 
\end{tablenotes}
\end{threeparttable}
\end{table}

\begin{table}
\centering
\caption{Estimated infectious units per million (IUPM) of HIV (with 95\% confidence intervals) for 17 individuals receiving care at the University of North Carolina HIV Cure Center\label{TableS5}}
\fontsize{10}{12}\selectfont
\begin{threeparttable}
\begin{tabular}{ccccc}
\toprule
\multicolumn{2}{c}{\textbf{ }} & \multicolumn{3}{c}{\textbf{Method}} \\
\cmidrule(l{3pt}r{3pt}){3-5}
\textbf{Subject} & \textbf{Bias} & \textbf{Without UDSA} & \textbf{With UDSA} & \textbf{With UDSA}\\
\textbf{ID} & \textbf{Correction} & \textbf{(Multiple Dilutions)} & \textbf{(Single Dilution)} & \textbf{(Multiple Dilutions)}\\
\midrule
C1 & MLE & $0.05$ $(0.02, 0.12)$ & $0.14$ $(0.08, 0.24)$ & $0.13$ $(0.07, 0.23)$\\
C1 & BC-MLE & $0.04$ $(0.02, 0.12)$ & $0.13$ $(0.08, 0.24)$ & $0.13$ $(0.07, 0.23)$\\
C2 & MLE & $0.07$ $(0.03, 0.15)$ & $0.17$ $(0.10, 0.29)$ & $0.18$ $(0.11, 0.29)$\\
C2 & BC-MLE & $0.07$ $(0.03, 0.15)$ & $0.17$ $(0.10, 0.29)$ & $0.18$ $(0.11, 0.29)$\\
C3 & MLE & $0.12$ $(0.05, 0.29)$ & $0.12$ $(0.05, 0.28)$ & $0.11$ $(0.04, 0.26)$\\
\addlinespace
C3 & BC-MLE & $0.12$ $(0.05, 0.29)$ & $0.11$ $(0.05, 0.28)$ & $0.10$ $(0.04, 0.26)$\\
C4 & MLE & $0.14$ $(0.06, 0.32)$ & $0.12$ $(0.05, 0.29)$ & $0.14$ $(0.06, 0.30)$\\
C4 & BC-MLE & $0.14$ $(0.06, 0.32)$ & $0.11$ $(0.04, 0.29)$ & $0.13$ $(0.05, 0.30)$\\
C5 & MLE & $0.15$ $(0.06, 0.36)$ & $0.20$ $(0.09, 0.44)$ & $0.20$ $(0.10, 0.43)$\\
C5 & BC-MLE & $0.14$ $(0.06, 0.36)$ & $0.18$ $(0.08, 0.44)$ & $0.19$ $(0.08, 0.43)$\\
\addlinespace
C6 & MLE & $0.18$ $(0.08, 0.38)$ & $0.34$ $(0.20, 0.60)$ & $0.31$ $(0.18, 0.54)$\\
C6 & BC-MLE & $0.17$ $(0.08, 0.37)$ & $0.33$ $(0.19, 0.59)$ & $0.30$ $(0.17, 0.53)$\\
C7 & MLE & $0.22$ $(0.13, 0.36)$ & $0.64$ $(0.49, 0.83)$ & $0.63$ $(0.48, 0.81)$\\
C7 & BC-MLE & $0.22$ $(0.13, 0.36)$ & $0.63$ $(0.48, 0.82)$ & $0.62$ $(0.48, 0.81)$\\
C8 & MLE & $0.41$ $(0.27, 0.62)$ & $1.06$ $(0.86, 1.31)$ & $1.06$ $(0.86, 1.30)$\\
\addlinespace
C8 & BC-MLE & $0.40$ $(0.26, 0.61)$ & $1.04$ $(0.84, 1.29)$ & $1.04$ $(0.84, 1.29)$\\
C9 & MLE & $0.44$ $(0.22, 0.87)$ & $0.60$ $(0.35, 1.03)$ & $0.51$ $(0.30, 0.88)$\\
C9 & BC-MLE & $0.42$ $(0.20, 0.86)$ & $0.56$ $(0.31, 1.00)$ & $0.48$ $(0.28, 0.85)$\\
C10 & MLE & $0.54$ $(0.32, 0.91)$ & $0.73$ $(0.48, 1.12)$ & $0.79$ $(0.53, 1.17)$\\
C10 & BC-MLE & $0.52$ $(0.30, 0.89)$ & $0.69$ $(0.44, 1.09)$ & $0.75$ $(0.50, 1.14)$\\
\addlinespace
C11 & MLE & $0.62$ $(0.34, 1.14)$ & $0.89$ $(0.55, 1.43)$ & $0.88$ $(0.57, 1.38)$\\
C11 & BC-MLE & $0.58$ $(0.30, 1.11)$ & $0.82$ $(0.49, 1.37)$ & $0.83$ $(0.51, 1.33)$\\
C12 & MLE & $0.94$ $(0.63, 1.39)$ & $1.42$ $(1.19, 1.69)$ & $1.42$ $(1.19, 1.69)$\\
C12 & BC-MLE & $0.90$ $(0.59, 1.36)$ & $1.40$ $(1.17, 1.67)$ & $1.40$ $(1.17, 1.67)$\\
C13 & MLE & $1.24$ $(0.74, 2.08)$ & $2.55$ $(1.22, 5.36)$ & $1.41$ $(0.79, 2.49)$\\
\addlinespace
C13 & BC-MLE & $1.12$ $(0.63, 1.99)$ & $2.32$ $(1.03, 5.25)$ & $1.19$ $(0.61, 2.34)$\\
C14 & MLE & $1.82$ $(0.90, 3.66)$ & $2.12$ $(0.86, 5.18)$ & $2.20$ $(1.12, 4.34)$\\
C14 & BC-MLE & $1.54$ $(0.67, 3.52)$ & $1.88$ $(0.69, 5.15)$ & $1.73$ $(0.73, 4.11)$\\
C15 & MLE & $2.49$ $(1.10, 5.62)$ & $3.66$ $(1.88, 7.10)$ & $2.95$ $(1.57, 5.57)$\\
C15 & BC-MLE & $2.08$ $(0.79, 5.52)$ & $3.27$ $(1.56, 6.87)$ & $2.62$ $(1.28, 5.36)$\\
\addlinespace
C16 & MLE & $2.52$ $(1.16, 5.46)$ & $2.92$ $(1.46, 5.84)$ & $2.97$ $(1.62, 5.44)$\\
C16 & BC-MLE & $2.18$ $(0.89, 5.33)$ & $2.65$ $(1.24, 5.69)$ & $2.70$ $(1.39, 5.26)$\\
C17 & MLE & $3.49$ $(1.69, 7.20)$ & $2.19$ $(0.98, 4.88)$ & $3.03$ $(1.66, 5.55)$\\
C17 & BC-MLE & $3.08$ $(1.36, 7.00)$ & $1.99$ $(0.82, 4.80)$ & $2.74$ $(1.40, 5.35)$\\
\bottomrule
\end{tabular}
\end{threeparttable}
\end{table}

\clearpage

\section{Derivations and Simulations for the Negative Binomial MLE}
\label{webE}

\subsection{Negative binomial likelihood}\label{webE1}

Suppose $X_{ij}$ follows a negative binomial distribution with mean parameter $\lambda_i$ and dispersion parameter $\gamma$ (shared across all DVLs), i.e., $X_{ij}$ has probability mass function (PMF)
\begin{align*}
    {\Pr}_{\lambda_i,\gamma}(X_{ij} = x_{ij}) =
    \frac{\Gamma(x_{ij} + 1/\gamma)}{\Gamma(1/\gamma)\Gamma(x_{ij} + 1)}
    \left(\frac{1}{\gamma \lambda_i + 1}\right) ^ {1/\gamma}
    \left(\frac{\gamma \lambda_i}{\gamma \lambda_i + 1}\right) ^ {x_{ij}}, 
\end{align*}
for $\gamma \in (0, \infty)$, $x_{ij} \in \{0, 1, 2, ...\}$, and $\Gamma(\cdot)$ denoting the gamma function \citep{agresti_categorical_2013}.
To allow $\gamma \in [0, \infty)$, define ${\Pr}_{\lambda_i,\gamma=0}(X_{ij} = x_{ij})$ as
\begin{align*}
    \lim_{\gamma \to 0}{\Pr}_{\lambda_i,\gamma}(X_{ij} = x_{ij}) =
    \frac{\exp(-\lambda_i)\lambda_i^{x_{ij}}}{x_{ij}!}.
\end{align*}
That is, if $\gamma=0$ then $X_{ij}$ follows a Poisson distribution with mean $\lambda_i$, as in Section 2 of the main text. When $\gamma>0$, ${\Pr}_{\lambda_i,\gamma}(X_{ij} = 0) = (\gamma \lambda_i + 1) ^ {-1/\gamma}$, so $Z_{ij}$ and $W_j$ follow Bernoulli distributions with $\Pr_{\lambda_i,\gamma}(Z_{ij} = 1) = 1 - (\gamma \lambda_i + 1) ^ {-1/\gamma}$ and $\Pr_{\pmb{\lambda},\gamma}(W_j = 1) = 1 - \prod_{i=1}^n (\gamma \lambda_i + 1) ^ {-1/\gamma}$. 
Following Sections 2.2 and 2.6 of the main text, when $\gamma>0$, the negative binomial observed-data likelihood for data from a single dilution level is proportional to $L_{NB}(\pmb{\lambda}, \gamma|M_N, \pmb{Y})=$  
\begin{align*}
    \prod_{i=1}^n \left[\left\{1 - (\gamma \lambda_i + 1) ^ {-1/\gamma}\right\}^{Y_i}
    (\gamma \lambda_i + 1) ^ {-(M_N + m - Y_i)/\gamma}\right]
    \left\{1 - \prod_{i=1}^n (\gamma \lambda_i + 1) ^ {-1/\gamma} \right\}^{M - M_N - m}, 
\end{align*}
and for data from multiple dilution levels the likelihood is proportional to
\begin{align}
\widetilde{L}_{NB} (\pmb{\tau}, \gamma|\pmb{M_N}, \pmb{Y}, \pmb{u}) =
\prod_{d=1}^D L_{NB}\left(u^{(d)}\pmb{\tau}, \gamma|M_{N}^{(d)}, \pmb{Y}^{(d)}\right).
\label{negbin-joint-likelihood}
\end{align}
To allow $\gamma \in [0, \infty)$, define $L_{NB}(\pmb{\lambda}, \gamma=0|M_N, \pmb{Y}) \equiv \lim_{\gamma \to 0}L_{NB}(\pmb{\lambda}, \gamma|M_N, \pmb{Y}) = L(\pmb{\lambda}|M_N, \pmb{Y})$ for the Poisson distribution in (1) of the main text.

Note that the negative binomial likelihood for a single dilution level $u$ depends on the parameters $\lambda_i=u\tau_i$ and $\gamma$ only through $\Pr_{\lambda_i,\gamma}(Z_{ij} = 1)$, and that different values of $\lambda_i$ and $\gamma$ can lead to the same $\Pr_{\lambda_i,\gamma}(Z_{ij} = 1)$. For example, $\Pr_{\lambda_i=1,\gamma=2}(Z_{ij} = 1) = \Pr_{\lambda_i=\sqrt{3}-1,\gamma=1}(Z_{ij} = 1) = 1-1/\sqrt{3}$. Thus, the negative binomial parameters are not identifiable given assay data from a single dilution level. On the other hand, it is straightforward to show that the negative binomial parameters \textit{are} identifiable given assay data from multiple dilution levels. In this case, the negative binomial MLE ($\btauhat_{NB}, \widehat{\gamma}_{NB})$ is obtained by maximizing the log of \eqref{negbin-joint-likelihood} over $\btau \in (0, \infty)^n$ and $\gamma \in [0, \infty)$ using 
the L-BFGS-B method \citep{ByrdEtAl1995}, a variation of BFGS that allows for the inequality constraint $\gamma \geq 0$. \citet{ByrdEtAl1995} suggest that the L-BFGS-B method will typically converge to a global maximum given a concave objective function.

\subsection{Likelihood ratio test for overdispersion}\label{webE2}

The presence of overdispersion can be assessed by testing the null hypothesis $H_0: \gamma = 0$ versus the alternative $H_1: \gamma > 0$. These hypotheses can be tested with a likelihood ratio test (LRT) using the test statistic $S_{LR} = 2 \log\{\widetilde{L}_{NB}
(\hat{\pmb{\tau}}_{NB}, \widehat{\gamma}_{NB}|\pmb{M_N}, \pmb{Y}, \pmb{u}) / \widetilde{L}(\hat{\pmb{\tau}}|\pmb{M_N}, \pmb{Y}, \pmb{u})\}$, comparing the maximum values of the Poisson and negative binomial likelihood functions. Since this is a test of one parameter ($\gamma$) with the null value ($0$) on the boundary of the parameter space, under $H_0$ the LRT statistic $S_{LR}$ has an asymptotic distribution that is a 50/50 mixture of $\chi^2_0$ and $\chi^2_1$ distributions \citep{self_asymptotic_1987}.

\subsection{Overdispersion simulations}\label{webE3}

The performance of the negative binomial MLE and the LRT test for overdispersion was investigated in a simulation study. Assay data were simulated following Section 3.2 for $D=3$ dilution levels. The following parameters were held fixed: (i) the true IUPM $T=1$, (ii) the three dilution levels $\pmb{u} = (u_1, u_2, u_3) = (0.5, 1, 2)$ million cells per well, (iii) the proportions of positive wells to be deep sequenced at the three dilution levels $\pmb{q} = (q_1, q_2, q_3) = (0, 0.5, 1)$, and (iv) $n'=12$ underlying DVLs. The simulation settings varied by the number of replicate wells per dilution level, $\pmb{M} = (M_1, M_2, M_3) = (6, 12, 18)$ or $(30, 60, 90)$, and the dispersion parameter $\gamma = 0$ (no overdispersion), $\gamma = 0.2$ (light overdispersion), $\gamma = 1$ (moderate overdispersion), and $\gamma = 4$ (heavy overdispersion). For each scenario, $500$ assays were simulated.

\begin{figure}
    \centering
    \includegraphics[width=\textwidth]{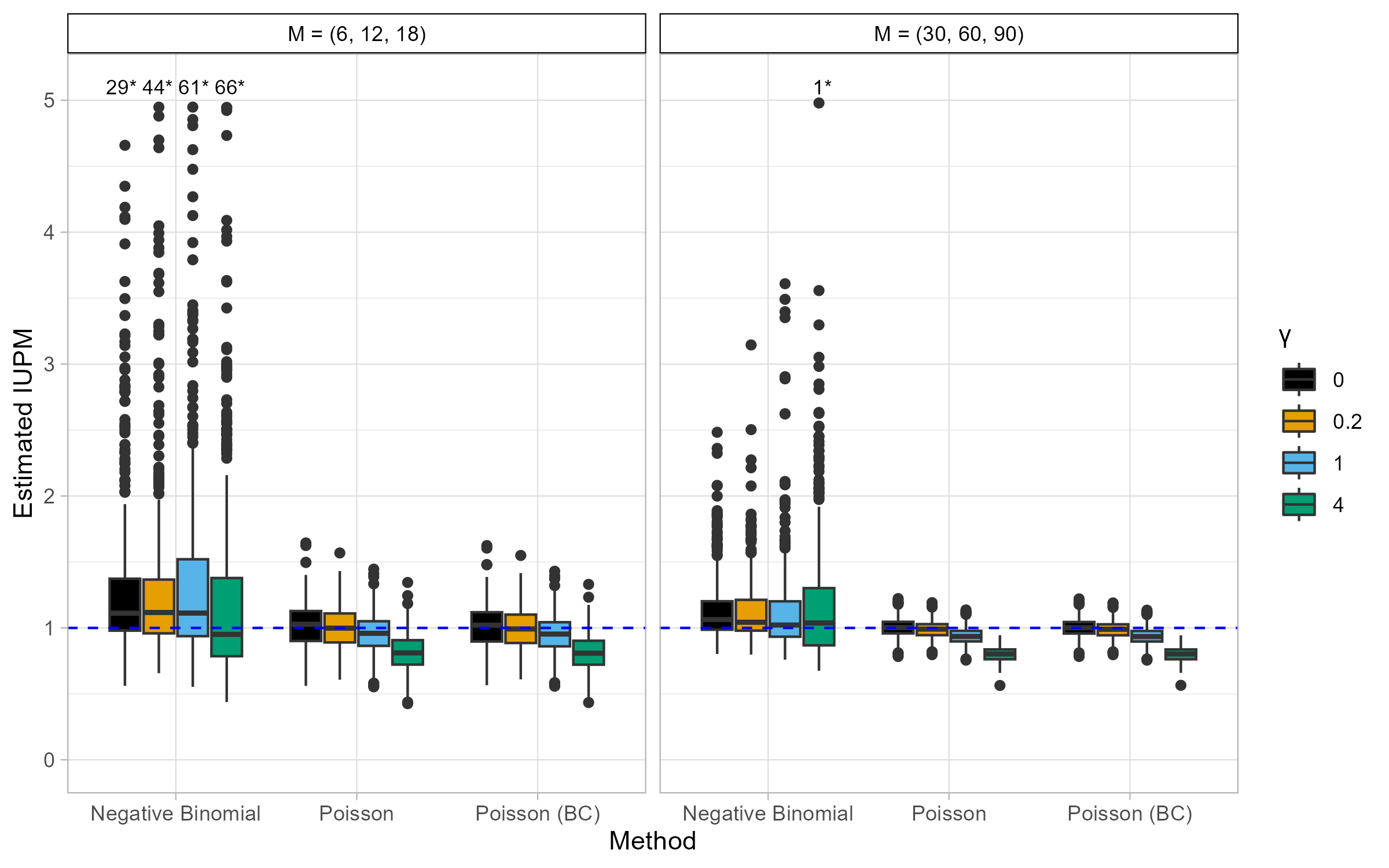}
    \caption{Empirical distributions of the estimated IUPM with $D=3$ dilution levels (0.5, 1, and 2 million cells per well), proportions of positive wells that were deep sequenced at the three dilution levels of $\pmb{q} =(0, 0.5, 1)$, $n'=12$ DVLs, IUPM of $T=1$, and \num{500} simulations per setting. *Number of values beyond the range of the plot.\label{FigureS1}}
    \label{fig:fig-s1}
\end{figure}

Figure~\ref{fig:fig-s1} shows the empirical distributions of three IUPM estimators under varying degrees of overdispersion: (i) the negative binomial MLE, (ii) the Poisson MLE, and (iii) the Poisson BC-MLE. As expected, when the counts of infected cells $X_{ij}$ follow a negative binomial distribution with nonzero $\gamma$, the Poisson MLE and Poisson BC-MLE are biased, regardless of the numbers of replicate wells $\pmb{M}$. For light to moderate overdispersion, this bias is relatively small, but it becomes severe under heavy overdispersion. The negative binomial MLE has notably smaller bias under overdispersion. However, in all scenarios, the variance of the negative binomial MLE is much larger than the variances of the Poisson MLE and BC-MLE. In terms of mean-squared error (MSE), a combined measure of bias and variance, the Poisson MLE and BC-MLE outperform the negative binomial MLE.

Table~\ref{table-lrt-power} shows the empirical power of the LRT of overdispersion at the 0.05 significance level against various alternatives. Under the null hypothesis, the LRT had approximately correct type 1 error rate ($\approx 0.05$) for both small and large numbers of replicate wells $\pmb{M}$. Under the alternative hypotheses, the power of the LRT increased as $\gamma$ increased (i.e., it became farther from the null value of $0$) and as the numbers of replicate wells $\pmb{M}$ increased. However, in all cases, the power of the LRT against overdispersion was very low.

\begin{table}
\centering
\caption{Empirical power of the likelihood ratio test for overdispersion\label{TableS6}}
\label{table-lrt-power}
\begin{threeparttable}
\begin{tabular}{ccc}
\toprule
\textbf{M} & $\pmb{\gamma}$ & \textbf{Power of LRT}\\
\midrule
(6, 12, 18) & 0.0 & 0.046\\
(6, 12, 18) & 0.2 & 0.064\\
(6, 12, 18) & 1.0 & 0.098\\
(6, 12, 18) & 4.0 & 0.104\\
\addlinespace
(30, 60, 90) & 0.0 & 0.042\\
(30, 60, 90) & 0.2 & 0.042\\
(30, 60, 90) & 1.0 & 0.082\\
(30, 60, 90) & 4.0 & 0.216\\
\bottomrule
\end{tabular}
\begin{tablenotes}[flushleft]
\footnotesize{\item{\em Note:} $\gamma$ is the negative binomial dispersion parameter; $\gamma = 0$ corresponds to the null hypothesis of no overdispersion; Power is the proportion of simulated replicates where the null hypothesis of no overdispersion was rejected at the 0.05 significance level.}
\end{tablenotes}
\end{threeparttable}
\end{table}

In summary, these simulation results demonstrate for the scenarios considered that it is generally difficult to detect overdispersion in the latent counts of infected cells given the observed well-level infection indicator variables. Moreover, despite correctly modeling overdispersion, the negative binomial MLE for the IUPM has much greater MSE than the Poisson MLE, regardless of the true dispersion parameter, due to its increased variance.

\subsection{Assessing overdispersion in the HIV application}\label{webE4}

The HIV data from Section 4 of the main text were assessed for overdispersion using the likelihood ratio test described in Web Appendix E.2. Table~\ref{table-lrt-real-data} shows the overdispersion LRT statistics and corresponding p-values, as well as the Poisson and negative binomial MLEs for each of the 17 participants.

\begin{table}
\centering
\caption{Overdispersion LRT results for HIV Application}\label{TableS7}
\label{table-lrt-real-data}
\begin{threeparttable}
\fontsize{10}{12}\selectfont
\centering
\fontsize{10}{12}\selectfont
\begin{tabular}{ccccc}
\toprule
\multicolumn{3}{c}{\textbf{ }} & \multicolumn{2}{c}{\textbf{Estimated IUPM}} \\
\cmidrule(l{3pt}r{3pt}){4-5}
\textbf{Subject ID} & \textbf{LRT Statistic} & \textbf{LRT P-Value} & \textbf{Poisson MLE} & \textbf{Negative Binomial MLE}\\
\midrule
C1 & 0.000 & 1.000 & 0.130 & 0.130\\
C2 & 0.174 & 0.338 & 0.178 & 0.315\\
C3 & 0.000 & 1.000 & 0.107 & 0.107\\
C4 & 0.412 & 0.261 & 0.135 & 0.350\\
C5 & 0.492 & 0.241 & 0.203 & 0.862\\
\addlinespace
C6 & 0.000 & 1.000 & 0.314 & 0.314\\
C7 & 0.000 & 1.000 & 0.627 & 0.627\\
C8 & 0.000 & 1.000 & 1.057 & 1.057\\
C9 & 0.000 & 1.000 & 0.515 & 0.515\\
C10 & 1.439 & 0.115 & 0.791 & 1.971\\
\addlinespace
C11 & 0.043 & 0.418 & 0.885 & 1.052\\
C12 & 0.010 & 0.460 & 1.418 & 1.510\\
C13 & 7.469 & 0.003 & 1.406 & 6.176\\
C14 & 0.000 & 1.000 & 2.203 & 2.203\\
C15 & 0.000 & 1.000 & 2.953 & 2.953\\
\addlinespace
C16 & 0.000 & 1.000 & 2.968 & 2.968\\
C17 & 0.483 & 0.244 & 3.034 & 4.679\\
\bottomrule
\end{tabular}
\end{threeparttable}
\end{table}

Aside from participant C13, the LRT tests provide no evidence of overdispersion and IUPM estimates are similar for the negative binomial and Poisson models. For participant C13 the LRT p-value is 0.003, suggesting possible overdispersion. However, this result should be interpreted with caution as the reported p-values are not adjusted for multiple testing. For example, if the LRT p-values for all subjects were compared to a 0.05 significance threshold, then the family-wise error rate would be $1-(1-0.05)^{17} \approx 0.58$. In addition, the results for subject C13 are sensitive to slight changes in the observed data, as demonstrated in Table~\ref{table-lrt-sens}.

\begin{table}
\centering
\caption{Sensitivity Analysis for Subject C13}\label{TableS8}
\label{table-lrt-sens}
\begin{threeparttable}
\centering
\fontsize{10}{12}\selectfont
\begin{tabular}{cccc}
\toprule
$\pmb{M_P}$  & \textbf{LRT Statistic} & \textbf{LRT P-Value} & \textbf{Negative Binomial MLE}\\
\midrule
(16, 4, 3, 0) & 7.469 & 0.003 & 6.179\\
(\underline{17}, 4, 3, 0) & 4.450 & 0.017 & 5.190\\
(\underline{18}, 4, 3, 0) & 0.057 & 0.406 & 3.361\\
(16, 4, \underline{2}, 0) & 5.058 & 0.012 & 4.204\\
(16, 4, \underline{1}, 0) & 3.145 & 0.038 & 2.937\\
\bottomrule
\end{tabular}
\begin{tablenotes}[flushleft]
\footnotesize{\item{\em Note:} The true counts of QVOA-positive wells for participant C13 is $\pmb{M_P}$ = (16, 4, 3, 0), which corresponds to the first row. Subsequent rows show the effect of changing one component of $\pmb{M_P}$ (underlined) by a small amount.}
\end{tablenotes}
\end{threeparttable}
\end{table}

\section{Extension for Imperfect Assay Sensitivity and Specificity}
\label{webF}

\subsection{Single dilution likelihood for imperfect assay sensitivity and specificity}\label{webF1}
For simplicity, focus on assay data from a single dilution level of one million cells per well. Let $W_j^*$ and $Z_{ij}^*$ denote  error-prone versions of the true infection indicators $W_j$ and $Z_{ij}$, respectively, obtained using imperfect assays. The true infection indicators $W_{j}$ and $\bZ_{j}$ are missing for all wells (i.e., they are latent variables). Instead, suppose that the imperfect QVOA results $\bW^* = (W_1^*, \dots, W_M^*)^\T$ are fully observed, while the imperfect UDSA results $\bZ^* = (\bZ_1^*, \dots, \bZ_M^*)^\T$ will have missing data for any unsequenced wells. Let $M_P^* = \sum_{j=1}^{M}W_{j}^*$ and $M_N^* = M - M_P^*$ denote the numbers of wells found to be HIV-positive and HIV-negative, respectively, using the imperfect QVOA, and let $m^* = \nint{qM_P^*}$ denote the number of positive wells to be deep-sequenced via the imperfect UDSA. Importantly, $W_j^* = 0$ no longer implies that $\bZ_j^* = \pmb{0}$, so the imperfect UDSA results are missing for all ($M - m^*$) unsequenced wells.

To incorporate latent $W_j$ and $\bZ_j$ in addition to partially missing $\bZ_j^*$, begin with a ``complete'' observation, $(R_j, W_j^*, \bZ_j^*, W_j, \bZ_j)$, assumed to be generated from the joint distribution 
\begin{align}
&\Pr(R_j, W_j^*, \bZ_j^*, W_j, \bZ_j) \nonumber \\
&= \Pr(R_j|W_j^*, \bZ_j^*, W_j, \bZ_j)\Pr(W_j^*|\bZ_j^*, W_j, \bZ_j)\Pr(\bZ_j^*|W_j, \bZ_j)\Pr(W_j|\bZ_j)\Pr(\bZ_j) \nonumber \\
& = \Pr(R_j|W_j^*)\Pr(W_j^*|\bZ_j^*, W_j, \bZ_j)\Pr(\bZ_j^*|W_j, \bZ_j)\Pr(W_j|\bZ_j)\Pr(\bZ_j), \label{joint_imperfect}
\end{align}
where $\Pr(R_j|W_j^*) = m^*/M_P^*$ by design; $\Pr(W_j^*|\bZ_j^*, W_j, \bZ_j)$ and $\Pr(\bZ_j^*|W_j, \bZ_j)$ are the error mechanisms of the imperfect QVOA and UDSA, respectively; and $\Pr(W_j|\bZ_j)\Pr(\bZ_j) = \Pr(W_j, \bZ_j)$ is the joint distribution of the true infection statuses from the perfect assays (discussed in Sections 2.1--2.6). Since the true QVOA result is fully determined by the true UDSA results, $\Pr(W_j|\bZ_j)\Pr(\bZ_j) = \Pr(\bZ_j)$ still.

Further assumptions are made about the error mechanisms. First, the error-prone assay results are assumed to depend only on the true assay results for the same test, such that the error mechanisms simplify to $\Pr(W_j^*|\bZ_j^*, W_j, \bZ_j) = \Pr(W_j^*|W_j)$ and $\Pr(\bZ_j^*|W_j, \bZ_j) = \Pr(\bZ_j^*|\bZ_j)$. Note that $\Pr(W_j^*=1|W_j=1)$ and $\Pr(Z_{ij}^*=1|Z_{ij}=1)$ are the sensitivities and $\Pr(W_j^*=0|W_j=0)$ and $\Pr(Z_{ij}^*=0|Z_{ij}=0)$ are specificites of the QVOA and UDSA assays, respectively. Second, the error-prone UDSA results for the $i$th DVL are assumed to depend only on the true UDSA results for the same DVL, such that $\Pr(Z_{ij}^*|\bZ_j) = \Pr(Z_{ij}^*|Z_{ij})$ and thus $\Pr(\bZ_j^*|\bZ_j) = \prod_{i=1}^{n}\Pr(Z_{ij}^*|Z_{ij}).$
Third, the sensitivity and specificity of both assays are assumed to be known (e.g., from the literature or a validation study). If sensitivity and specificity values are not known, then analyses may be repeated over a range of plausible values to assess robustness of the IUPM estimates against imperfect assays.

With these assumptions, the joint distribution simplifies from \eqref{joint_imperfect} to
\begin{align}
& \Pr(R_j|W_j^*)\Pr(W_j^*|W_j)\Pr(\bZ_j^*|\bZ_j)\Pr(\bZ_j), 
\label{joint_imperfect_simplified}
\end{align}
where $\Pr(\bZ_j) = \prod_{i=1}^{n}\Pr(Z_{ij})$ following from the assumption of independence between the counts of infected cells across DVLs, as in the main text.

To incorporate all available information on all wells, the observed-data likelihood function for imperfect assays can be constructed from \eqref{joint_imperfect_simplified}, and is proportional to
\begin{align}
L(\blambda|\bW^*,\bZ^*,\bR) &= \prod_{j=1}^{M}{\Pr}_{\blambda}(W_j^*,\bZ_j^*)^{R_j}{\Pr}_{\blambda}(W_j^*)^{1-R_j}, \label{likelihood-imperfect}
\end{align}
where $\Pr_{\blambda}(W_j^*,\bZ_{j}^*)$ is the PMF of the imperfect assay results $(W_j^*, \bZ_j^*)$ and $\Pr_{\blambda}(W_j^*)$ is the marginal PMF of the imperfect QVOA result $W_j^*$. Notice that \eqref{likelihood-imperfect} follows the same general form as $L(\blambda|\bW,\bZ,\bR)$ 
under perfect sensitivity and specificity from the main text, except that now $\bW^*$ are fully observed (instead of $\bW$), while $\bZ^*$ has some missing data (instead of $\bZ$). The imperfect UDSA results $\bZ^*$ are still MAR for the $M - m^*$ unsequenced wells, so $\Pr(R_j|W_j^*)$ can be omitted from the likelihood for $\blambda$ \citep{Little&Rubin2002}.

The PMFs for the imperfect assay results depend on the true assay results and can be calculated by marginalizing out the unobserved variables to obtain 
\begin{align}
{\Pr}_{\blambda}(W_j^*,\bZ_j^*) & = \sum_{z_1=0}^{1}\cdots\sum_{z_n=0}^{1}\prod_{i=1}^{n}\Pr(W_j^*|W_j=w)\Pr(Z_{ij}^*|Z_{ij}=z_i){\Pr}_{\lambda_i}(Z_{ij}=z_i), \label{PW*Z*} 
\end{align}
for $w = \I(\sum_{i=1}^{n}z_i \geq 1)$, and 
\begin{align}
{\Pr}_{\blambda}(W_j^*) & = \sum_{w = 0}^{1}\Pr(W_j^*|W_j=w){\Pr}_{\blambda}(W_j=w), \label{PW*}
\end{align}
where, as in the main text, ${\Pr}_{\lambda_i}(Z_{ij}) = \left\{1-\exp(-\lambda_i)\right\}^{Z_{ij}}\exp(-\lambda_i)^{(1-Z_{ij})}$ and ${\Pr}_{\blambda}(W_j) = \left\{1 - \exp\left(-\sum_{i=1}^{n}\lambda_i\right)\right\}^{W_j}\exp\left(-\sum_{i=1}^{n}\lambda_i\right)^{1-W_j}$. Note that when there is no assay error, \eqref{PW*Z*} and \eqref{PW*} simplify to ${\Pr}_{\blambda}(W_j, \bZ_j)$ and ${\Pr}_{\blambda}(W_j)$, respectively, from Section 2.2.

When the assays have imperfect sensitivity and specificity, the MLE $\lambdahat_i$ for detected DVL $i$ is not guaranteed to be positive. In this case, the L-BFGS-B method introduced in \ref{webE} is used to impose the inequality constraint $\lambda_i \geq 0$. As in \ref{webA}, supplying the gradient of the log-likelihood based on \eqref{likelihood-imperfect} improves the performance of the numerical optimization procedure. The elements of this gradient are given by
\begin{align*}
\frac{\partial}{\partial\lambda_i}l(\blambda|\bW^*,\bZ^*,\bR) 
 &= \sum_{j=1}^{M}R_j\frac{\frac{\partial}{\partial\lambda_i}{\Pr}_{\blambda}(W_j^*,\bZ_j^*)}{{\Pr}_{\blambda}(W_j^*,\bZ_j^*)} + \sum_{j=1}^{M}(1-R_j)\frac{{\frac{\partial}{\partial\lambda_i}\Pr}_{\blambda}(W_j^*)}{{\Pr}_{\blambda}(W_j^*)},
\end{align*}
where $\frac{\partial}{\partial\lambda_i}{\Pr}_{\blambda}(W_j^*,\bZ_j^*) =$
\begin{align*}
&  \sum_{z_1=0}^{1}\cdots\sum_{z_n=0}^{1}\Pr(W_j^*|W_j=w)\Pr(\bZ_{j}^*|\bZ_{j}=\bz)\left\{\prod_{i'\neq i}{\Pr}_{\lambda_{i'}}(Z_{i'j}=z_{i'})\right\}\left\{(-1)^{1-z_{i}}\exp(-\lambda_i) \right\}
\end{align*}
and $\frac{\partial}{\partial\lambda_i}{\Pr}_{\blambda}(W_j^*)  = $
\begin{align*}
& \sum_{w = 0}^{1}\Pr(W_j^*|W_j=w)(-1)^{1-w}\exp\left(-\sum_{i'=1}^{n}\lambda_{i'}\right).
\end{align*}

\subsection{Multiple dilution likelihood for imperfect assay sensitivity and specificity}\label{webF2}

Modifications to \ref{webF1} for different and multiple dilutions are straightforward. First, consider assay data from a single dilution level of $u \times 10^6$ cells per well, for some $u>0$. The imperfect assays likelihood at this dilution level is proportional to $\widetilde{L}(\btau|\bW^*,\bZ^*,\bR, u) = L(u\btau|\bW^*,\bZ^*,\bR)$, following the same $\btau = \blambda/u$ substitution from Section 2.6 in the main text. Similarly, the imperfect assays likelihood for $D$ distinct dilution levels, $D \in \{1, 2, \dots \}$, is proportional to
\begin{align}
\prod_{d=1}^D \widetilde{L}(\btau|\bW^{*(d)},\bZ^{*(d)},\bR^{(d)}, u^{(d)}),\tag{S.22}
\label{multiple-likelihood-imperfect}
\end{align}
where $\bW^{*(d)}$, $\bZ^{*(d)}$, $\bR^{(d)}$, and $u^{(d)}$ denote the vector of imperfect QVOA results, matrix of imperfect UDSA results, vector of complete data indicators, and dilution level, respectively, for the $d$th dilution. As in Section 2.6 the imperfect assays MLE for the vector of DVL-specific IUPMs is the value $\btauhat$ that maximizes \eqref{multiple-likelihood-imperfect}, and the corresponding MLE for the IUPM can be calculated as $\Tauhat=\sum_{i=1}^n\tauhat_i$.

\subsection{Simulations for imperfect assay sensitivity and specificity}\label{webF3}

The performance of the imperfect assays MLE was investigated in a simulation study. Following the steps outlined in Section 3.1, true assay results $W$ and $\bZ$ were simulated for $M = 12, 24$, or 32 replicate wells at a single dilution level. A dilution of $u = 1$ million cells per well was chosen with a true IUPM of $1$, such that $T = \Lambda = 1$. The IUPM was allocated at a constant rate across $n' = 6$ underlying DVLs, such that $\lambda_i = 1/6$, $i \in \{1, \dots, 6\}$. For comparison, the ```perfect assays'' IUPM estimator from the main text was included, which assumed 100\% sensitivity and specificity for both the QVOA and UDSA.

Three sets of simulations were conducted at a single dilution level. In the first set, combinations of QVOA and UDSA sensitivities between $80\%$, $90\%$, and $100\%$ were considered, with specificity fixed at $90\%$ for both assays. In the second set, the assays' specificities were varied while their sensitivity was fixed at $90\%$ and $q = 1$. In the first two sets of simulations, all wells found to be HIV-positive using the imperfect QVOA would undergo deep sequencing (i.e., $q = 1$). In the third set, the proportion of wells testing positive that underwent deep sequencing varied between $q = 0.5, 0.75$, and $1$. In these simulations, the QVOA and UDSA had 90\% sensitivity and 90\% specificity.

\begin{figure}
    \centering
    \includegraphics[width=\textwidth]{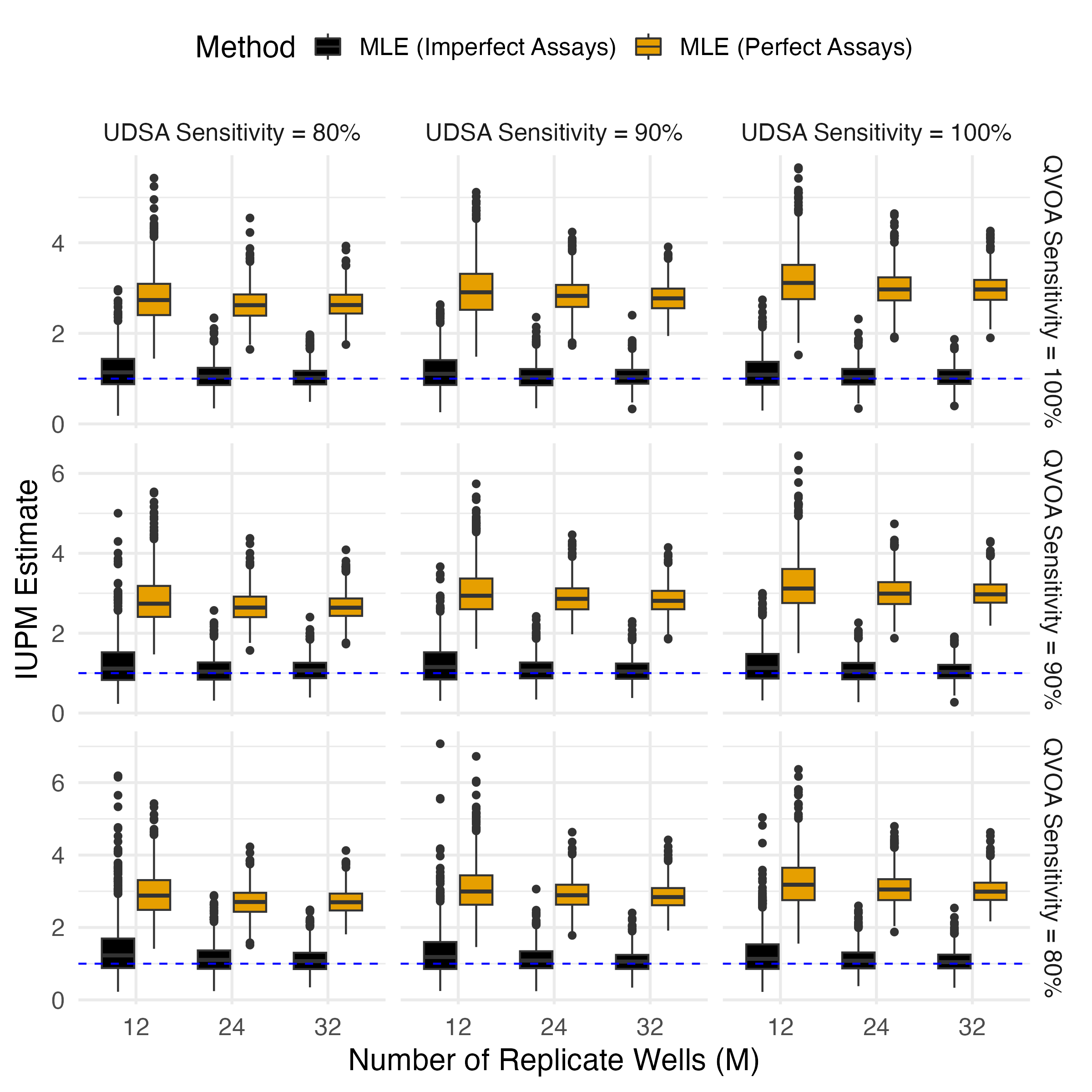}
    \caption{Empirical distributions of the imperfect assays MLE and the perfect assays MLE of the IUPM at a single dilution level. QVOA and UDSA sensitivities varied, but both assays had 90\% specificity. Two replicates where the imperfect assays MLE was $> 10$ were excluded from the plot.\label{FigureS2}}
    \label{fig:fig-s2}
\end{figure}

\begin{figure}
    \centering
    \includegraphics[width=\textwidth]{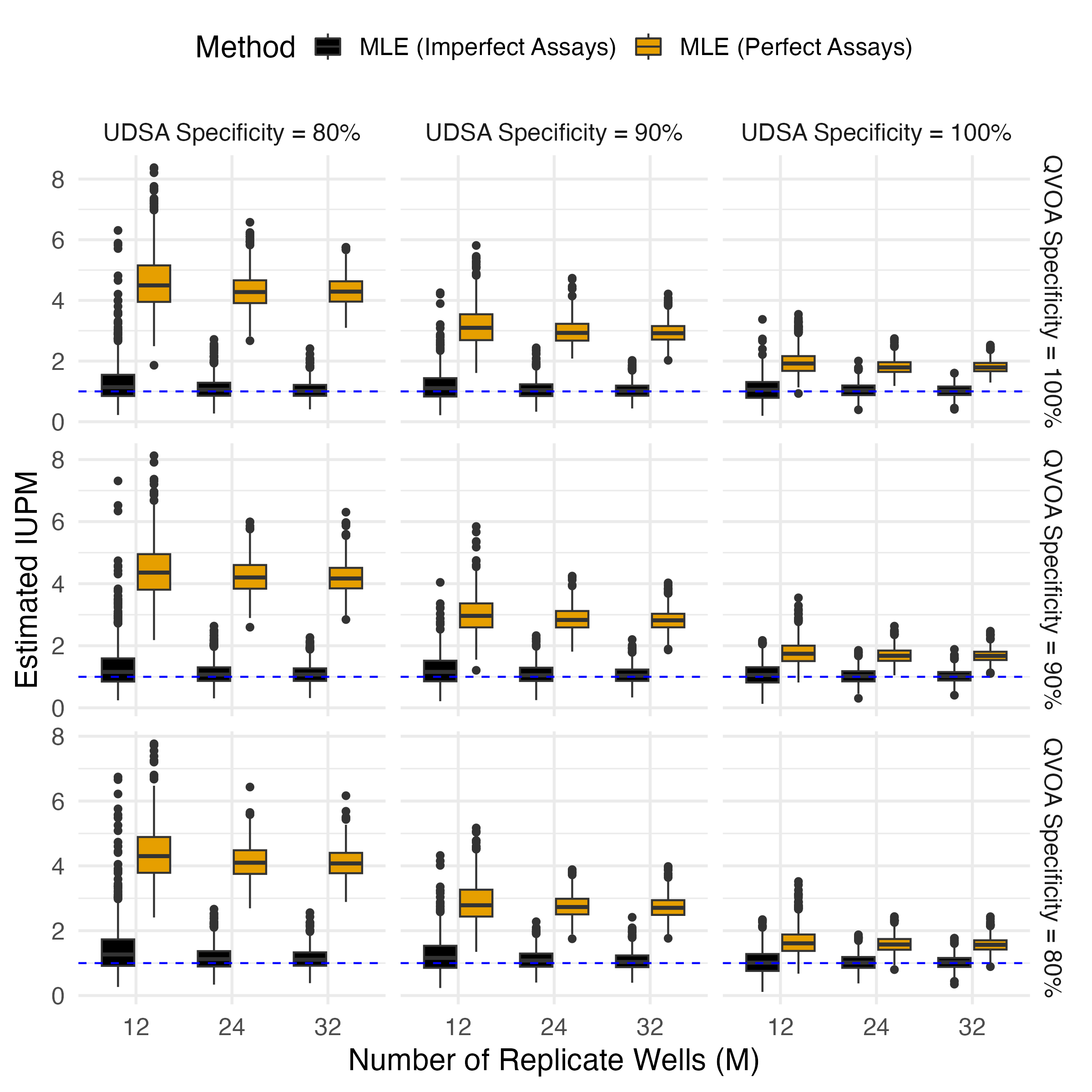}
    \caption{Empirical distributions of the imperfect assays MLE and the perfect assays MLE of the IUPM at a single dilution level. QVOA and UDSA specificities varied, but both assays had 90\% sensitivity. One replicate where the imperfect assays MLE was $> 10$ was excluded from the plot.\label{FigureS3}} 
    \label{fig:fig-s3}
\end{figure}

\begin{figure}
    \centering
    \includegraphics[width=\textwidth]{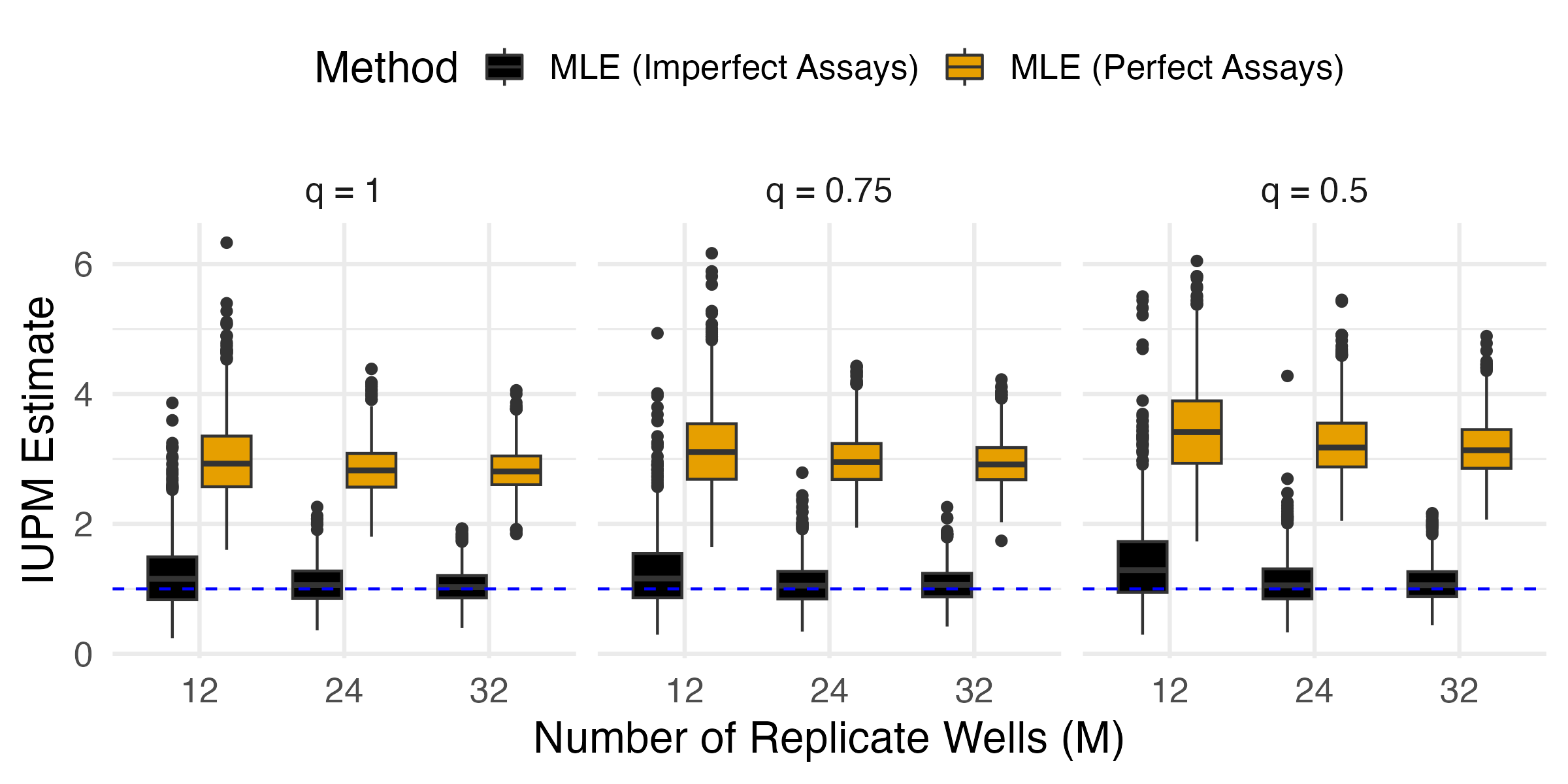}
    \caption{Empirical distributions of the imperfect assays MLE and the perfect assays MLE of the IUPM at a single dilution level. The proportion of imperfectly HIV-positive wells to undergo deep sequencing varied, but both assays had 90\% sensitivity and 90\% specificity. One replicate where the imperfect assays MLE was $> 10$ was excluded from the plot.\label{FigureS4}}
    \label{fig:fig-s4}
\end{figure}

Figures~\ref{fig:fig-s2} and \ref{fig:fig-s3} show the empirical distributions of the imperfect assays MLE and the perfect assays MLE of the IUPM under varying assay sensitivities and specificities, respectively. As expected, the perfect assays MLE was biased when the perfect sensitivity and specificity assumption was violated. This bias increased as the sensitivity and specificity decreased and persisted as the number of replicate wells $M$ increased. Across all settings considered, the imperfect assays MLE had very little bias, even as the amount of missing deep sequencing data increased (Figure~\ref{fig:fig-s4}).

\begin{figure}
    \centering
    \includegraphics[width=\textwidth]{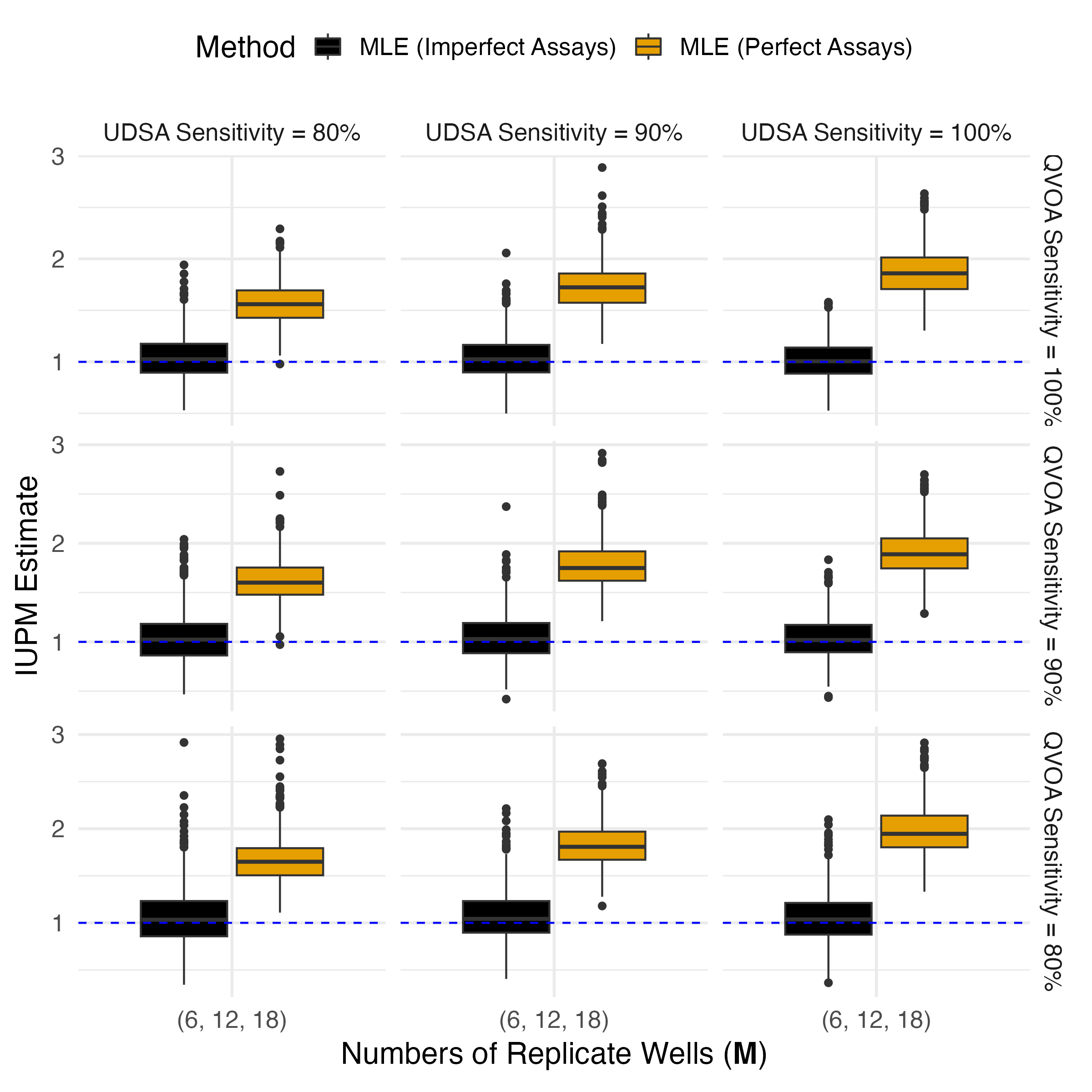}
    \caption{Empirical distributions of the imperfect assays MLE and the perfect assays MLE of the IUPM at multiple dilution levels. QVOA and UDSA sensitivities varied, but both assays had 90\% specificity. One replicate where the imperfect assays MLE did not converge was excluded from the plot.\label{FigureS5}}
    \label{fig:fig-s5}
\end{figure}

An additional set of simulations was conducted at multiple dilution levels. Following the steps outlined in Section 3.2, true assay results at $D = 3$ dilution levels were simulated with $\pmb{M} = (M_1, M_2, M_3) = (6, 12, 18)$ replicate wells per dilution level. The true IUPM of $1$ was constantly distributed across $n' = 6$ underlying DVLs, such that $\lambda_i = 1/6$, $i \in \{1, \dots, 6\}$. For the imperfect assay results, combinations of QVOA and UDSA sensitivities between $80\%$, $90\%$, and $100\%$ were considered, with specificity fixed at $90\%$ for both assays. The imperfect assays MLE continued to perform well for assay data from multiple dilution levels, with very little bias across various error settings (Figure~\ref{fig:fig-s5}).  

\subsection{Considering imperfect assays in the HIV application}\label{webF4}

\begin{figure}
    \centering
    \includegraphics[width=\textwidth]{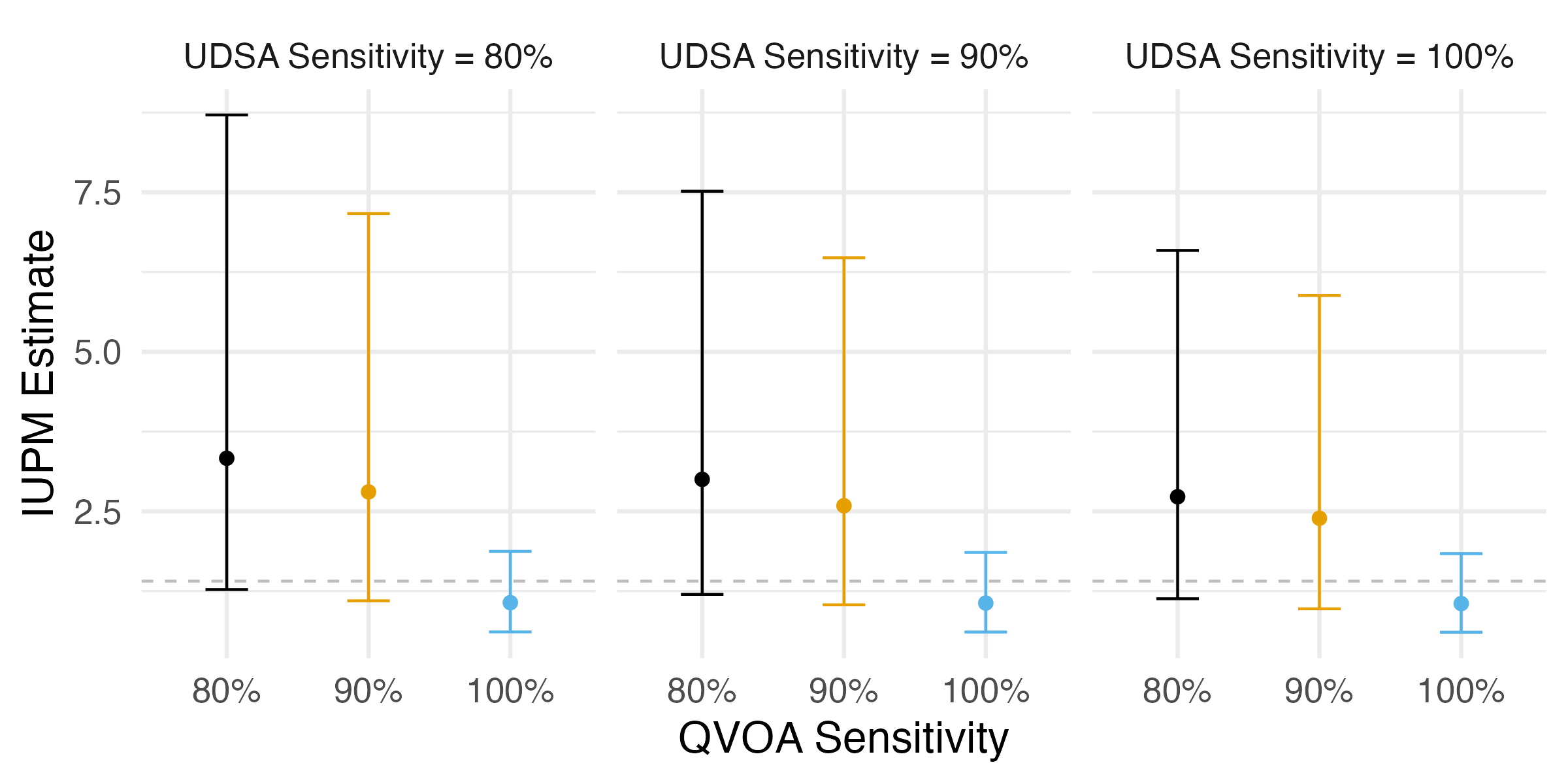}
    \caption{Estimated IUPM with 95\% confidence intervals for Subject C13, assuming imperfect assays. QVOA and UDSA sensitivities varied, but both assays were assumed to have 90\% specificity. The dashed line denotes the estimated IUPM assuming perfect assays.\label{FigureS6}}
\end{figure}

The HIV data from Section 4 of the main text were used to demonstrate how imperfect assay sensitivity and specificity can be considered in practice. Specifically, the IUPM estimates of Subject C13 were computed under 90\% specificity and varied sensitivity (the same settings considered for the simulations in Figure~\ref{fig:fig-s5}) and compared to the IUPM estimates assuming perfect sensitivity. Note the methods in \ref{webF3} above require per-well data (i.e., $W_{j}^*$ and $Z_{ij}^*$). Unfortunately, the QVOA and UDSA results were summarized by the lab so that such per-well level data was not available. Therefore, well-level data consistent with the observed summary data were simulated in order to illustrate application of the methods in \ref{webF3} The resulting IUPM estimates and 95\% confidence intervals for Subject C13 are shown in Figure~\ref{FigureS6}. As as either the UDSA or QVOA sensitivity increased from 80\% to 100\%, the IUPM estimates decreased and the corresponding confidence intervals became narrower. This process could be repeated for all subjects (i.e., source populations).

\label{lastpage}

\clearpage
\bibliographystyle{biom} 
\bibliography{mybib.bib}